\documentstyle[12pt]{article}
\newcommand{\eq}[1]{(\ref{#1})}
\newcommand{\diff}{\partial}
\newcommand{\beq}{\begin{equation}}
\newcommand{\eeq}{\end{equation}}
\newcommand{\beqn}{\begin{eqnarray}}
\newcommand{\eeqn}{\end{eqnarray}}
\newcommand{\cD}{{\cal D}}

\def\cZ{{\cal Z}}

\def\cR{{\cal R}}

\def\cC{{\cal C}}
\def\cW{{\cal W}}

\def\NP{ Nucl.~Phys.}
\def\PR{ Phys.~Rev.}
\def\PL{ Phys.~Lett.}
\def\PRL{ Phys.~Rev.~Lett.}

\begin{document}

\hfill {\bf ITEP-TH-46/97}

\hfill hep-th/9711153
\vspace{10mm}

\centerline{\bf \Large Black Hole Thermodynamics}
\centerline{\bf \Large from the point of view of Superstring
Theory}

\vspace{5mm}

\centerline{E.T.~Akhmedov\footnote{e--mail: akhmedov@vxitep.itep.ru}}

\centerline{Institute of Theoretical and Experimental Physics}

\centerline{Moscow, 117259, B. Cheremushkinskaya, 25.}

\vspace{10mm}

\begin{abstract}
In this review we try to give a pedagogical introduction to the
recent progress in the resolution of old problems of black hole
thermodynamics within superstring theory. We start with a brief description
of classical black hole dynamics. Then, follow with the consideration
of general properties of supesymmetric black holes. We conclude with the
review of the statistical explanation of the black hole entropy and string
theory description of the black hole evaporation.
\end{abstract}




\section*{\bf 1.Introduction}

   Being just classical solutions in General Relativity, black holes
\cite{MiThWh73} behave as if they possess entropy \cite{Bek73} and
temperature \cite{Haw74}.  It was argued that this fact causes some
puzzles which we describe in the subsections 1.1,
1.2, 1.3. They can not be resolved, at least naively, if gravity is
considered as a classical background for quantum theory
\cite{Bek73,Haw74,Haw76,Prob}. Then, quantum gravity is required.  We
believe that any reasonable, i.e.  self-consistent, quantum theory of
gravity would not have these puzzles \cite{tHo96}.  At present we have at
our disposal only one such a theory which is that of {\it
superstrings} \cite{GrScWi}. Therefore, a good exercise is to check if
superstring theory really has no puzzles under consideration.

  From the modern point of view General Relativity and low energy string
theory live at different points of some space of parameters of fundamental
{\it M-theory} \cite{rewdu,WiHuTo95} which is yet to be formulated
\cite{BaSu}.  In fact, one gets General Relativity from the fundamental
theory when all scales in the latter are big in comparison with the string
one $\sqrt{\alpha'}$. So that all string theory corrections to the General
Relativity are small.  While the string gas description is valid when the
string coupling constant $g_s$ tends to zero.

   Although one still has to work to get a phenomenologically
reasonable theory, that of superstrings is considered as the most
promising {\bf self consistent} theory which quantizes gravity and,
even more, unifies it with Yang-Mills fields. As we discuss in this
paper, superstring theory also provides an explanation of the black
hole thermodynamics.  However, our today's knowledge about this
theory is limited.  Therefore, our answers also are not quite
complete.

    Let us proceed now with the brief discussion of the puzzles and of
the progress we are going to review.

\section*{\bf 1.1. Entropy}

   Black holes in General Relativity serve as {\it attractors}
-- objects present in the configuration spaces of some
non-linear dynamical systems.  We mean that independently of initial
conditions before a collapse, there is a static solution characterized by a
few parameters, after it, as $t\to\infty$. In other words, the static black
hole solutions do not "remember" initial conditions.  This fact manifests
itself in the so called {\it No Hair theorem} \cite{NoHair}. It establishes
that one can characterize the {\bf static} black hole solutions only by their
masses, angular momenta and different charges corresponding to local
symmetries. Qualitatively, argumentation goes as follows \cite{NoHair}:
first, semiclassically a remote observer can see nothing behind an event
horizon; second, gravity is sensitive only to the
energy-momentum tensor; third, the frequency of quanta, emitted by an object
falling into the black hole, tends to zero when measured by a distant
observer \cite{LaLi}.  Hence, after the collapse, as $t\to\infty$, one can
feel different internal black hole charges, corresponding to local
symmetries, only via static fields.

   All this means that the static black holes possess a degeneracy
over initial values of any parameters (like multi-pole moments,
global quantum numbers and etc.) in a theory containing General Relativity.
Exceptions are those which appear in the statement of the No Hair theorem.
Manifestation of this degeneracy is the appearance of the entropy $S$
which is attributed to the black holes themselves \cite{Bek73}. It is
proportional to the area $A$ of the event horizon \cite{Bek73}:  $S =
A / 4\Gamma_N$, where $\hbar=c=1$ and $\Gamma_N$ is the Newton
constant.

   This entropy was calculated via the use of the laws of classical
black hole dynamics only \cite{Bek73,Haw74}. {\bf Thus, one needs to
find a statistical explanation of the entropy.} In principle to solve the
problem one can calculate the degeneracy corresponding to the internal states
of a black hole. This way one finds \cite{NoFr} the proportionality $S\sim
A$.  If this is not satisfactory for some reasons, for example, because of
the possibility of the information loss \cite{Haw76}, one might consider
fluctuations of the event horizon \cite{tHo96,Horiz}. Keeping the black hole
mass and charges fixed, we find that the number of such "fluctuation
states" is proportional to exponent of the horizon area.  Hence, again
one can qualitatively recover the proportionality\footnote{In three
dimensions one even can recover the proper coefficient of the proportionality
\cite{Car94}.} $S\sim A$ \cite{Horiz}.

   {\bf Thus, the real question is how to explain the coefficient of this
proportionality.} For example, in the just described reasoning, the
coefficient is divergent.  That is why, the correct statistical derivation of
the entropy should include counting of the black hole internal
states in quantum gravity. In this case the black hole degeneracy can be
lifted.

   As we have already said, the most promising candidate for
quantum theory of gravity is that of superstrings. Within this theory
the black holes of General Relativity correspond to some quantum
string states. In fact, when the string coupling constant $g_s$ tends to
zero, the black hole horizon shrinks \cite{Sus93,HoPo96} and becomes smaller
than the string scale $\sqrt{\alpha'}$.  At
this point the geometric description by General Relativity is not appropriate
and one has a pure string state.  Thus, we can calculate the degeneracy of
the latter and compare its log with the black hole horizon area.

  Why should one expect an agreement between these two numbers?
To answer on this question we consider the four-dimensional Schwarzschild
black hole as in ref. \cite{HoPo96}.  Its mass is equal to $M_{bh} = r_0 /
2\Gamma_N$, where $r_0$ is the radius of the event horizon.  We want to
equate the latter with the mass of a string state at the excitation level
$N$, which is $M^2_s\sim N / \alpha'$ at zero string coupling $g_s$.  The
Newton constant is related to the string coupling $g_s$ and $\alpha'$ by
$\Gamma_N\sim g^2_s\alpha'$.  So it is clear that the mass of the black hole
cannot be equal to the string mass for all values of $g_s$.  If we want to
equate them, we have to decide at what value of the string coupling they
should be equal.  Clearly, the natural choice is to let $g_s$ be the value at
which the string forms a black hole, which is when the horizon is of the
order of the string scale $\sqrt{\alpha'}$. Setting the masses $M^2_{bh}$ and
$M^2_s$ equal when $r_0^2\sim \alpha'$ yields $\alpha'/ \Gamma_N^2 \sim N /
\alpha'$, which happens when $g_s\sim N^{-\frac14}$.  The black hole entropy
is then $S_{bh} \sim r_0^2 / \Gamma_N \sim \alpha'/ \Gamma_N \sim \sqrt{N}$.
At the same time, because the string state is a composition of the oscillator
states its degeneracy could be evaluated as follows.  It is equal to the
number of splittings $N = \sum_j j n_j$ of $N$ into the numbers $j$, each
appearing $n_j$ times in the sum.  Simple combinatorical
exercise\footnote{One considers the generation function of those numbers
\cite{GrScWi}:  $G(\omega) = tr\left(\omega^{\hat{N}}\right) =
\sum^{\infty}_{n=0} d_n \omega^n$, where $\hat{N} = \sum^{\infty}_m
a^{+}_m a_m$ is the operator of number of particles.  We are looking
for the number $d_{N}$. At the same time $tr(\omega^{\hat{N}}) =
\prod^{\infty}_{m=1} tr (\omega^{a^+_m a_m}) =
\prod^{\infty}_{m=1}(1-\omega^m) = \exp{\left(- \sum^{\infty}_{m=1}
ln(1-\omega^m)\right)} = \exp{\left( \sum^{\infty}_{m,n=1}
\frac{\omega^{mn}}{n}\right)}\sim \exp{\left(\frac{1}{1 - \omega}
\sum_{m=1}^{\infty}\frac{1}{m^2}\right)} = \exp{\left(
\frac{\pi^2}{6(1-\omega)}\right)}$. Therefore, $d_{N} =
\frac{1}{2\pi i} \int \frac{G(\omega) d\omega}{\omega^{N+1}} \sim
\exp{(const\cdot\sqrt{N})}$.} shows that in the large $N$ limit
the degeneracy is proportional to $e^{\sqrt{N}}$. Thus, we see a
qualitative agreement between $S_{bh}$ and $S_{s}$ for the black hole.
This qualitative reasoning can be generalized to any dimension and
to charged black holes \cite{HoPo96}.

  To find an agreement between $S_{bh}$ and $S_s$ up to the
factor $1 / 4\Gamma_N$, we need some special circumstances
\cite{YoVo97}. For not much is yet known about string theory. The
circumstances are as follows. Theorists use such black holes which
have the entropy independent of the string coupling
constant. There are solitons of this kind in superstring theory.  Being {\it
supersymmetric (SUSY)} charged extreme black holes, they saturate
so called {\it Bogomolni-Prasad-Sommerfield (BPS)} bound.  These
solitons survive quantum corrections because they belong to some
specific representations of the SUSY algebra.  In fact, during a
smooth variation of the coupling constants in a theory
representations of symmetry algebras can not change \cite{WiOl79}.
Among the BPS black holes one should consider those, which are regular on
the event horizon. Because singular event horizons lead to that string
theory corrections for the event horizon area become strong.

   That is why theorists are forced to use some kind of a tricky
tuning of parameters of the black hole solutions in the
{\it supergravity (SUGRA)} theories. Therefore, following \cite{Mal96}, we
consider a black hole which corresponds to the bound state of {\it D-branes}
\cite{Pol}, as $g_s \to 0$. The latter are string states described as
manifolds on which strings can terminate.  The black hole in
question carries both electric and magnetic charges under several
electromagnetic like gauge fields and obeys the BPS bound.  The presence of
several charges helps to have a BPS solution with singularities of all
fields, defining the solution, shifted from the event horizon. {\bf As we
review below, in this case string theory perfectly explains the black hole
entropy including the numerical factor} $1 / 4\Gamma_N$.

   One comment is in order at this point. The extreme black holes
are fictitious from the thermodynamic point of view.  While having
non-zero entropy, they have zero temperature and, hence, do not
evaporate.  Therefore, we continue by considering slightly non-extreme black
holes.

\section*{\bf 1.2. Evaporation}

   The entropy, discussed in the previous subsection, is a response
function on variations of the temperature. It happens because
the black holes behave as if they have the temperature \cite{Haw74}. In
this subsection we try to give a qualitative explanation of this
fact.

   The black hole evaporation is just an ordinary decay process of
a state in quantum theory combined with quantum tunneling. In fact,
the black holes radiate because there can be a pair creation in the
strong gravitational field \cite{DeW75}.  During such a process one
of the created particles falls down the black hole while the other
escapes to infinity.  This should be the case due to the energy
conservation law: one of the particles acquires negative while the
other equal positive energy.

  Whether for the evaporation in the stationary situation\footnote{For an
evaporation in the non-stationary situation the presence of the event
horizon is not necessary \cite{DeW75}. However, in this case
the evaporation is not thermal.} one should have a black hole or just
a very massive object, can be checked as follows.
The first (very qualitative) argument is:  the energy of two created
particles should be equal to the modulus of the bending energy of one
of them with the evaporating body. We mean that $\Gamma^{(4)}_N
\frac{Mm_1}{r_1} = m_1 + m_2$, where $M$ is the mass of the
evaporating body, $m_{1,2}$ are masses of the created particles and
$r_1$ is the radius at which the particle creation is possible.
Thus, the radius $r_1$ is always smaller than the Schwarzschild one
which is equal to $\Gamma^{(4)}_N M$.  Hence, in the stationary
situation only the objects possessing event horizons can evaporate.

   The second argument goes as follows: for such an evaporation to
occur, space-time should be geodesic incomplete \cite{Haw94}, hence
singular. Therefore, it should contain an event horizon\footnote{One
comment is in order here.  There might be non-singular but
geodesically incomplete space-times.  For example, one can consider a
collapse of a matter with pressure rather than that of the cosmic
dust.  However, this is non-stationary situation which evaporates
non-thermally and eventually will tern into some static black
hole.}.  In fact, {\it geodesic incompleteness} means, almost by
definition, that there might be particles whose history has a
beginning or end at a finite proper time.  Thus, particle creation
might happen only in the presence of a black hole \cite{Haw94}.

   It was argued \cite{Haw76,Haw94} that due to the presence of the event
horizon one should average over all "invisible" states behind its surface.
Doing this, one gets a mixed quantum state.  However, this is not a whole
story.  The main fact about black holes, as we have already mentioned, is
that one gets the thermal mixed state. For the static black hole
solutions this thermal nature can be shown by considering quantum
field theory in their background.  After the Wick rotation from the
Minkovski space to the Euclidean one, we should transform to the
coordinates in which the metric is regular on the event horizon. This way one
gets a well defined theory in which the time coordinate is necessarily
an angular variable \cite{Haw94}.  Thus, obtained theory is thermal,
with the length of the compact time direction being inverse temperature.  Due
to the No Hair theorem we have a finite number of the static solutions and
can check the thermal behavior of all of them.  However, these considerations
are semiclassical \cite{calcul} and applicability of the semiclassics in this
case in arguable \cite{tHo96}.

  {\bf Thus, without knowledge of quantum gravity, one can have
the following "puzzle" \cite{Haw76,Haw94}:  before a collapse there is
a pure quantum state, after it, there is the mixed thermal one.} Being
a non-unitary transition it is a forbidden process in quantum theory.
To resolve this puzzle we should find some unitary description of the
black hole radiation. We believe that in any quantum theory of
gravity the radiation is described by a unitary process \cite{tHo96}.
However, now at our disposal we have only superstring theory.
Hence, if it pretends to be a quantum gravity it should give such a
process.

   Therefore, in this paper we study, following ref.  \cite{MaSt96,DhMaWa96},
the evaporation of slightly non-extreme black holes. First we consider the
process in the black hole and then in the D-brane pictures.  For the
calculation of the decay rate in the first case, one can use, for
example, quantum tunneling method.  While in the second case, it is
necessary to find a process which is a counterpart of the black hole
radiation.  For this reason one should consider some non-BPS excitations of a
D-brane bound state.  In our case, these excitations are given by a gas of
non-BPS open (attached to the D-branes) strings.  Two such strings can
collide to form a closed one, escaping to infinity.  Via standard quantum
theory methods one can calculate the amplitude of this process.  The
radiation is thermal due to the canonical distribution of the strings in
the gas \cite{DhMaWa96}.  {\bf As we review below, when one has a slight
deviation from the extremity, there is a perfect agreement between the decay
rates under consideration.} Therefore, we know the unitary string theory
description of the black hole radiation.

\section*{\bf 1.3. Information loss}

   We have already seen that in the semiclassical approximation one naively
looses information about states felt inside a black hole. This was the reason
for the averaging over intrinsic states. One can pose this "puzzle" in
another way \cite{Haw94}.  We shell skip all obvious assumptions in the
description below.

  Suppose that we have created somehow a pair of extreme black
holes. After that, one of them can absorb a particle, carrying some
conserved global charge, and evaporate back to the
extremity.  In the semiclassics the No Hair theorem establishes that one can
not test such a global charge inside any static black hole.  Moreover, the
{\bf thermal} evaporation does not carry any information about the
black hole intrinsic state.  Therefore, it seems that after such a process,
the pair of black holes can annihilate back, loosing information about the
quantum number under consideration.  {\bf Hence, one can argue that there is
an information loss in the black hole presence.}

   This is not quite true, however. For in the above consideration
we have implicitly supposed that the {\bf semiclassical} No  Hair
theorem is valid also on the quantum level.  It is even possible to
give arguments \cite{Haw94} in favor of this fact. Without
possibility to check such an information loss process experimentally,
this assumption might be reasonable if one could construct
self-consistent {\bf non-unitary} quantum theory. However, nobody yet
was able to construct such a theory.  At the same time, there is a
self-consistent unitary theory which gives a quantum description of the black
holes.  They are given by D-brane bound states for some of them. In any case,
in superstring theory one can measure the quantum state of any black hole.
Hence, within superstring theory, in the described above process black holes
will annihilate {\bf if and only if} they are equivalent. It happens if and
only if information is taken away by the radiation.

   One may argue, keeping in mind that black hole and D-brane
pictures are valid at different values of string theory coupling
constants, that our arguments are wrong \cite{Str96}.  In fact, a neutron
star has no event horizon and, hence, does not contradict unitarity.  While
if one will slightly change the value of the Newton constant, the neutron
star would collapse to form a black hole which can contradict unitarity.

    However, the above mentioned process can be considered as a
quantum mechanics on the two black hole {\it moduli space}
\cite{Str96}. The latter is the space of, invisible at infinity, parameters
defining a solution. In our case these are just black holes positions. The
main point here is that the topology of the moduli spaces does
not change during smooth variations of coupling constants \cite{Str96}. At the
same time, the moduli space of a D-brane bound state is the product
of their individual moduli spaces. This space should be
factored over the permutation group, acting on the D-brane positions,
{\bf if and only if} they are equivalent.  Therefore, the same is
true in the black hole case and our above arguments are correct.
{\bf This fact shows that there is no the information loss "paradox"
within superstring theory.}

\section*{\bf 1.4. Content of the review}

  Our review is organized as follows. In the second section we
briefly review the classical black hole dynamics. The third section includes
a discussion of the BPS states in low dimensional SUSY theories, for
simplicity of the presentation.  In this context we describe properties of
the extreme SUSY black holes in four- and five- dimensional SUGRA theories.

  In the section four we show what kind of the ten-dimensional SUGRA
solutions correspond to the black holes in dimensions smaller than ten.
After that we argue which string excitations should quantize
certain charged black hole solutions of the SUGRA theories.

   In the section five we calculate thermodynamic quantities for the extreme
black holes which are regular on the event horizon. We find
that they have zero temperature and non-zero entropy.  Then in this section
we calculate the degeneracy of the quantum string excitations corresponding
to these black holes.  Logs of the obtained numbers perfectly coincide
with the black hole areas.  Also in the section five we continue with the
review of the results on the radiation of the particular non-extreme black
holes. Here we also find a perfect agreement between the string theory and
black hole calculations.

   Conclusions are given in the section six. Also to make the review
self-contained as much as possible we included the discussion of some
basics of superstring theory and of the superstring duality in the
Appendices.

\section*{\bf 2. Black Hole Physics}

    In this section we present some basic facts concerning classical
General Relativity solutions themselves and their dynamics.

\section*{\bf 2.1. Black Hole solutions}

    The simplest black hole solution is that due to Schwarzschild in four
dimensions \cite{MiThWh73}. It is spherically symmetric solution of
the free Einstein-Hilbert equations with asymptotically flat
boundary conditions. It corresponds to the line element:

\beqn
ds^2 = - \Bigl(1 - \frac{2M'}{r}\Bigr) dt^2 + \Bigl(1 -
\frac{2M'}{r}\Bigr)^{-1} dr^2 + r^2 d\Omega^2_2,
\nonumber\\ d\Omega^2_2 = d\theta^2 + sin^2(\theta) d\phi^2,
\label{Sch}
\eeqn
where $M' = \Gamma^{(4)}_N M$ and $M$ is the black hole mass measured
by asymptotics of the energy-momentum tensor at infinity.

   The Schwarzschild solution has the {\it future event horizon}
at $r=2M'$. Semiclassically one can not see anything behind this
surface.  In the highly simplified context of spherically symmetric
and static space-time geometries the definition of the event horizon
is as follows\footnote{While for general space-times there are
many subtleties involved \cite{WaHaEl73}.}. This is a {\it null
surface}, i.e. the surface to which a normal vector has zero norm,
from behind of which a particle can not escape to infinity without
exceeding speed of light.

   The solution \eq{Sch} has also the {\it past} event horizon
which is the time reverse of the first one:  a surface which is
impossible to get behind. However, one can construct a black hole
solution which appeared as the result of a {\it collapse} and, hence,
has no past event horizon \cite{WaHaEl73}.

   As can be seen, the metric \eq{Sch} has the singularity at the
surface $r=2M'$.  It is referred to as a {\it coordinate singularity}
because the invariant of the curvature tensor is regular there.
Hence, the singularity can be avoided by some coordinate change.  Also there
is a {\it curvature singularity} at the point $r=0$, which can not be avoided
by any coordinate change. In fact, the invariant of the curvature tensor is
singular there.

  There are also rotating (Kerr), charged (Reissner-Nordstrom)
and both charged and rotating (Kerr-Newman) black hole solutions in
General Relativity \cite{MiThWh73}.  Below we will not discuss
rotating black holes. But all methods we use can be also applied in
the latter case \cite{Vaf96}.

   The Reissner-Nordstrom black hole is spherically symmetric
solution of the Einstein-Hilbert plus Maxwell equations of motion
with asymptotically flat boundary conditions.  In four dimensions the
metric of the solution looks as follows:

\beqn
ds^2 = - \Delta dt^2 + \Delta^{-1} dr^2 + r^2 d\Omega^2_2,
\nonumber\\ \Delta = \Bigl(1 - \frac{r_+}{r}\Bigr)\Bigl(1 -
\frac{r_-}{r}\Bigr), \quad r_+\ge r_-. \label{ReNo}
\eeqn
In proper units its mass and charge are:

\beqn
M = \frac{1}{2 \Gamma^{(4)}_N}(r_+ + r_-), \qquad Q =
\frac{1}{\Gamma^{(4)}_N}\sqrt{r_+r_-}. \label{mc}
\eeqn
The Reissner-Nordstrom black hole has the coordinate singularity at
the {\it outer} (event) horizon $r=r_+$. After the
definition of quantum theory in the black hole background, the
regularity of the energy-momentum tensor quantum average on
the event horizon is required.  Which leads to the energy
momentum tensor divergence at $r=r_-$ and produces the curvature
singularity at the {\it inner} horizon $r=r_-$.  Also, as can be seen
from \eq{mc}, always $M \ge Q$ \cite{Pen79}.  Solutions with $M=Q$
are referred to as {\it critical} or {\it extreme}.  They have
peculiar features, discussed in the following sections.

\section*{2.2. Black hole thermodynamics}

There are two main laws of classical black hole dynamics
\cite{Bek73,Haw71}.  The first one can be derived, for example, from
the expression for the black hole horizon area

\beq
A = 4\pi\Gamma^{(4)}_N \left[2 M^2 - Q^2 +
2M\sqrt{M^2 - Q^2}\right] \label{are}
\eeq
through its charge and mass from eq. \eq{mc}. One gets that variations of
these quantities are releted to each other as:

\beqn
dM = \frac{\kappa}{8\pi\Gamma^{(4)}_N} dA + \Phi dQ, \label{term}
\eeqn
where $\kappa = \frac{2\pi(r_+ - r_-)}{A}$ is referred to as
{\it surface gravity} and $\Phi = \frac{4\pi Q r_+}{A}$ is the
electromagnetic potential of the event horizon.

   The second law establishes that {\bf the area of a black hole event
horizon never decreases} \cite{Haw71}. In fact, as can be shown \cite{Haw94},
the null geodesic segments, which generate the event horizon, can not be
converging.

   At this point one can grasp an analogy with the first and second
laws of thermodynamics, respectively! This might be thought just as a
coincidence. But in a moment we will see that black holes really
behave as thermodynamic objects.

  Let us explain how the Schawrzschild black hole radiates
\cite{DeW75,Haw74}. We gave already a qualitative explanation of the black
hole radiation in the Introduction.  Let us give another one in this
subsection \cite{Haw74,calcul}.

   One should define quantum field theory in the curved background of
a black hole which was created by the collapse of a matter. Question
of prime importance is what one should consider as vacuum.  For this reason,
one needs to find a time direction to define what we mean by the energy
itself, by the lowest energy state and by asymptotic states.
However, the metric of a black hole created by a collapse is time
dependent.  Hence, there are different time-like {\it Killing
vectors} (that which define time direction) in different parts
of space-time.  Therefore, in different parts of space-time one has
different vacua.  Transformations between them are of the Bogolubov
type, i.e. as that in the Bardeen-Couper-Shrifer theory of
superconductors.  These transformations lead to the generation of
matter from geometry. The mixed state appears after an averaging over
the particles gone behind the horizon.  This way one gets the
following formula for the Schwarzschild black hole decay rate (only
due to a neutral particle radiation):

\beqn
d\Gamma = \sigma_{gb}(k_0)
\left(\frac{1}{\exp{\frac{2\pi k_0}{\kappa}} - 1}\right)
\frac{d^3k}{(2\pi)^3}. \label{emiss}
\eeqn
Here $k_0$ and $k$ are the energy and wave vector of an
escaping particle, $\kappa = \frac{1}{8\pi\Gamma^{(4)}_N M}$ and
$\sigma_{gb}$ is the so called {\it graybody factor}. It is equal
to the absorption cross section of the black hole\footnote{We discuss
$\sigma_{gb}$ for a particular solution in the subsection 5.3.}.
Looking at the formula \eq{emiss}, we can recognize the thermal
behavior of the black hole, with $T = \frac{\kappa}{2\pi}$ being its
temperature. Moreover, comparing \eq{term} (when $Q=0$) with the
second law of thermodynamics when $T = \frac{\kappa}{2\pi}$, we can
equate the entropy with $\frac{1}{4\Gamma^{(4)}_N} A$.

   Due to this thermal radiation, the black holes loose their entropy
and mass.  The first fact leads to the generalized second law of
thermodynamics: {\bf "entropy of a matter outside a black hole plus
$\frac{1}{4 \Gamma^{(4)}_N}$ times the area of the black hole event
horizon never decreases"}.  While because of the second fact the
Schwarzschild black hole will evaporate until $M\to 0$. Size of such
a limiting black hole is of the order of the Plank scale and there
will be strong quantum gravitational effects.

   The decay rate for the Reissner-Nordstrom black hole is the same
as in \eq{emiss} but with:

\beq
T = \frac{(r_+ - r_-)}{A}. \label{kappa}
\eeq
Therefore, if one considers only neutral particle emission, the
Reissner-Nordstrom black hole will evaporate until $T=0$ when $M=Q$
($r_+=r_-$). The end point of such an evaporation is a stable
extreme black hole (massless black hole is also considered as an
extreme $M=Q=0$) with the minimal allowed mass as discussed
below eq. \eq{mc}. In the next section we will see how SUSY explains
this stability.

\section*{3. Supersymmetric black holes}

  The main tool which gives to superstring theory control over the low
energy dynamics and over the dynamics of the extreme black holes
is the SUSY \cite{WeBa} in the target space.  The reason why one can
control the dynamics is that SUSY gives strong restrictions on the
effective low energy action. This is due to the {\it
non-renormalization theorems} \cite{GrRoSi79}. Also among all the
excitations of the SUSY theories there are remarkable ones. They are
referred to as BPS states.  We will discuss them first in the context of the
black hole physics below.  In this section we explain what are these BPS
states and their relation to the extreme black holes.

   Below we consider only $N\ge 2$ SUSY and SUGRA theories. In this case
one has almost complete control over the low energy dynamics.  We start
with the discussion of the BPS states. Then we consider peculiar features of
the four- and five-dimensional extreme black holes in the $N\ge 2$ SUGRA
theories.

   The $d$-dimensional $N$ extended SUSY algebra looks as follows
\cite{WeBa}:

\beqn
\left\{Q^I_{\alpha}, Q^J_{\beta}\right\} = \delta^{IJ}
\left(\cC\gamma^{\mu}\right)_{\alpha\beta} P_{\mu} +
\sum_{p=0,1,...}\left(\cC\gamma^{\mu_1...\mu_p}\right)_{\alpha\beta}
\cZ^{IJ}_{\mu_1...\mu_p}. \label{susy}
\eeqn
Here $Q^I_{\alpha}$, $I = 1,...,N$, being Dirac fermions, are
generators of the SUSY algebra and $\alpha$ are spinor indexes;
$P_{\mu}$ is the $d$-momentum; $\cC$ is the charge conjugation
operator and $\gamma^{\mu_1...\mu_p}$ is the anti-symmetric product
of $p$ $d$-dimensional Dirac $\gamma$-matrices;
$\cZ^{IJ}_{\mu_1...\mu_p}$ are so called {\it central charges} or {\it
central extensions}.  Not all of them are independent. For example, all pairs
$\cZ^{IJ}_{\mu_1...\mu_p}$ and $\cZ^{IJ}_{\mu_1...\mu_{d-p}}$ are
related to each other through the absolutely anti-symmetric tensor in
$d$-dimensions.

   Now, as an example let us show how one can derive BPS bound for a
particle state. We will consider the case of the four-dimensional $N=2$
SUSY algebra.  In the Weyl representation for the SUSY generators
this algebra looks as follows

\beqn
\left\{Q^I_{\alpha}, \bar{Q}^J_{\dot{\beta}}\right\} =
\sigma^{\mu}_{\alpha\dot{\beta}} P_{\mu}\delta^{IJ}, \nonumber\\
\left\{Q^I_{\alpha}, Q^J_{\beta}\right\} =
\epsilon_{\alpha\beta} \cZ^{IJ}, \qquad
\left\{\bar{Q}^I_{\dot{\alpha}}, \bar{Q}^J_{\dot{\beta}}\right\} =
\epsilon_{\dot{\alpha}\dot{\beta}} \cZ^{IJ}.
\eeqn
Therefore, from the positive semi-definiteness of the
anti-commutator of $\left(\sigma^{\mu}_{\alpha\dot{\beta}}
P_{\mu} Q^I_{\alpha} - \cZ\bar{Q}^I_{\dot{\beta}}\right)$
and its hermitian conjugate
$\left(\sigma^{\mu}_{\alpha\dot{\beta}}P_{\mu}
\bar{Q}^J_{\dot{\beta}} - \bar{\cZ} Q^J_{\alpha}\right)$
the BPS inequality follows \cite{WiOl79}:

\beq
- P_{\mu}^2 = M^2 \ge |\cZ|^2,
\eeq
where $M$ is the mass of a particle state and $\cZ^{IJ} =
\epsilon^{IJ}\cZ$ is the central charge in the corresponding
representation of the $N=2$ SUSY algebra. The central charge is
equal to a linear combination of the electric and magnetic charges of
the particle state \cite{WiOl79}.

  The BPS bound is saturated on the BPS states $M^2 = |\cZ|^2$. This
happens when the equality \cite{WiOl79}
$\left(\sigma^{\mu}_{\alpha\dot{\beta}} P_{\mu}
Q^I_{\alpha} - \cZ\bar{Q}^I_{\dot{\beta}}\right) |BPS> = 0$
holds.  Hence, the BPS states obey kind of a "chirality" condition.
Vise versa, if some state is annihilated by any linear combination of
the supercharges, it saturates the BPS bound. As a result of this
chirality less amount of terms is present in the Taylor expansion
of the corresponding superfield \cite{WeBa} over the anticommuting
variables.  That is why such states compose so called {\it ultra-short}
representation of the SUSY algebra.

   Now we come to a crucial point. It is believed that there are no
quantum corrections which can change the representation of any
algebra. Surely if the symmetry under this algebra persists on the quantum
level.  In the case of SUSY it means that {\bf if some state is BPS
semiclassically, it is BPS at any value of different coupling
constants of the theory!} Therefore, such a state survives quantum
corrections. One also can control the renormalization of its mass
due to the equality $M=|\cZ|$.  Moreover, if there is enough SUSY, this
mass does not even renormalize. As well, a BPS state is stable if
the process of its decay into lighter BPS states is forbidden by the
energy conservation law.  Hence, we can calculate the dimension of
the corresponding representation at any convenient value of the
couplings in a theory. One can be sure that such a calculation is true for
any other their values.  Similar fact in superstring theory
\cite{Sen95} will be important below.

   The generalization of all that to the $N=4$ and $N=8$ SUSY is
straightforward. Now the central charge $\cZ^{IJ}$ is a
matrix. Therefore, one can have different kinds of BPS states.  A
state composes the ultra-short multiplet if its mass is equal to all
eigen-values of the central charge.  While if the mass of a state is equal
only to the biggest eigen-value(s) of the central charge, then the state is
in a {\it short} BPS multiplet.  The latter break more than half of the
SUSY transformations. Hence, they are bigger than the ultra-short
one.

   In the presence of gravity, certain BPS solutions of the classical
equations of motion in SUGRA theories have line elements of the
extreme black holes.  In this case, the BPS condition is
equivalent to the extremity one.  Thus, stability of the BPS states
explains somewhat "magic" stability of the critical black holes.

   Now let us briefly discuss, following \cite{FeKaSt95}, some properties of
the black hole entropy in the $N\ge 2$ SUGRA theories. In the presence of the
$N\ge 2$ SUSY, one might expect that the black hole entropy would be
expressed (in addition to their mass, charge and momentum) through the {\it
vacuum expectation values (vev)} of different scalar fields in the
theory\footnote{One refers to these scalar fields as {\it moduli}. Their vevs
parametrize the {\it vacuum moduli space} or {\it flat directions} in the
theory. In the $N\ge 2$ SUSY theories, containing Yang-Mills fields, there
are scalar potentials of the following type:

\beq
V(\phi) = \frac{1}{2} tr \left([\phi^+, \phi]\right)^2. \label{pot'}
\eeq
They appear after the integration over the D-fields \cite{WeBa}. Here
$\phi$, being the superpartner of a gauge field, is the moduli type
field.  The potential under consideration vanishes when the $\phi$ field
lives in the maximal diagonal or abelian subalgebra of the gauge
symmetry algebra.  Hence, arbitrary complex number, standing on the diagonal
of the field $\phi$, parametrizes the moduli space of vacua.}.  However, the
BPS black holes do not depend on them.  The scaler fields behave as {\it
attractors} in the space of moduli.  Starting from their value at infinity,
moduli evolve, approaching the event horizon, until they run into a fixed
point near the horizon.  Therefore, by the time moduli reach the horizon they
lose completely the information about the initial conditions.  This is a kind
of a generalization of the No Hair theorem for the SUSY BPS black holes.

   The explanation of this phenomenon comes from the following
fact. The black hole line element interpolates between the flat
geometry at infinity and so called {\it Bertotti-Robinson (BR)}
geometry at the horizon. As it appears, even if the BPS black hole
solution preserves only some part of SUSY transformations, these two
boundary solutions (with corresponding gauge fields) preserve
all of it \cite{Gib85}. This restricts the boundary values of
the moduli fields.

   Briefly, it works as follows. The explicit form of the BR metric
is taken as a limit near the horizon $r\to 0$ of the black hole
metric:

\beqn
ds^2 = -h^2dt^2 + h^{-2}d\bar{r}^2, \nonumber\\
\Delta h^{-1} = 0, \label{geom}
\eeqn
where: $h^{-2} = \frac{A}{4\pi\Gamma^{(4)}_N r^2} =
\frac{M^2_{BR}}{\Gamma^{(4)} r^2}$ and the BR mass is defined
by the black hole area of the horizon:

\beq
M^2_{BR} = \frac{A}{4\pi}.\label{brmass}
\eeq
It can be straightforwardly checked that the geometry \eq{geom},
with corresponding gauge fields\footnote{Expressions for these
gauge fields will be given below for particular solutions.},
preserves all SUSY \cite{FeKaSt95,Gib85}.  From the formula
\eq{brmass} one can find the expression for the entropy through the
corresponding central charge which is equal to the BPS $M_{BR}$ mass.
Hence, the area of the horizon is proportional to the square of the
central charge of the SUSY algebra. The latter is computed at the point where
it is extremized in the moduli space \cite{FeKaSt95}. Hence,

\beq
S=\frac{A}{4\Gamma^{(4)}_N} = \pi |\cZ_{fix}|^2 \label{brent}
\eeq
in four dimensions. Similar calculation gives \cite{FeKaSt95}:

\beq
S=\frac{A}{4\Gamma^{(5)}_N} \sim |\cZ_{fix}|^{\frac32}
\eeq
for five-dimensional BPS black holes.

  Computing the extreme value of the central charge, one finds that
the entropy is expressed through some ratio of the black hole
charges. Thus, it is some kind of topological quantity:  it depends
only upon discrete parameters in the theory.  This observation will
appear useful for the discussion in the section five.

\section*{4. SUGRA solitons}

  In the previous section we have discussed the BPS solutions in the
low dimensional SUGRA. Among them there are non-perturbative
excitations.  At the same time, from its definition \cite{GrScWi}
(see also the Appendix A) string theory might be thought as
intrinsically perturbative. In fact, it is defined as the sum over
perturbative corrections.

   This is not quite true, actually, because there do exist
non-perturbative excitations in the superstring theory.  Now we are
ready to discuss variants of these fluctuations which are among
massive SUGRA solitons.  Concretely, this section is devoted to the
general discussion of how one constructs the classical solutions of
the ten-dimensional Type II SUGRA equations of motion \cite{DuKh94}.
Then we consider compactifications of the ten-dimensional SUGRA to
diverse dimensions. This way one can get particle-like black
holes from these SUGRA solitons. At the end we follow with the consideration
of the quantum counterparts of some black holes in superstring theory.

  Below we use only Type II superstring theories \cite{GrScWi}.
Let us sketch what are their massless excitations. For the
details one can consult the Appendix B. For our porpoises we need to
know only bosonic excitations of these theories.  They are subdivided
into two types. First type is that of $NS-\tilde{NS}$, which is
common for both of the Type II theories.  Moreover, it contains the
same fields as in the bosonic string theory. These are the graviton
$G_{\mu\nu}$, antisymmetric tensor $B_{\mu\nu}$ and dilaton
$\varphi$. Also one can add the dual or {\it "magnetic"} variant of
the $B_{\mu\nu}$ field in the following ten-dimensional sense:

\beq
H_{\mu_1...\mu_7} = e^{-2\varphi} \epsilon_{\mu_1...\mu_{10}}
H^{\mu_8\mu_9\mu_{10}},
\eeq
where $H_{\mu_1\mu_2\mu_3}$ is the field strength of the
antisymmetric tensor field $B$ and $H_{\mu_1...\mu_7}$ is that of the
dual field with six indexes.

    There is also $R-\tilde{R}$ sector.  Fields from this sector in
the Type IIA and IIB theories are different. We present them
separately for both of these theories.

{\bf In the Type IIA theory:}\\
$A_{\mu}$ is the 1-form potential; $A_{\mu\nu\alpha}$ is the 3-form
potential. Also there are dual, "magnetic", fields:

\beqn
F_{\mu_1...\mu_8} = \epsilon_{\mu_1...\mu_{10}}
F^{\mu_9\mu_{10}} \nonumber\\
F_{\mu_1...\mu_6} = \epsilon_{\mu_1...\mu_{10}}
F^{\mu_7...\mu_{10}}. \label{sd}
\eeqn
where $F$ are the field strengths of the corresponding gauge
potentials $A$.

    Superstring theory contains at low energies that of SUGRA \cite{GrScWi}.
For example, the bosonic part of the low energy SUGRA action of the Type IIA
theory looks as follows:

\beqn
S_{IIA} = \frac{1}{16\pi\Gamma^{(10)}_N}\int d^{10}
x\sqrt{-G}\Bigl\{e^{-2\varphi}\left[\cR +
4\left(\nabla_{\mu}\varphi\right)^2 - \frac13 H^2_{(3)} \right] -
\nonumber\\ - \alpha' F^2_{(2)} - \frac{\alpha'}{12} F'^2_{(4)} -
\frac{\alpha'}{288}\epsilon^{\mu_1...\mu_{10}}F_{\mu_1...\mu_4}
F_{\mu_5...\mu_8}B_{\mu_9\mu_{10}}\Bigr\} + O(\alpha'), \label{2a}
\eeqn
Here $\Gamma^{(10)}_N = 8\pi^6 g_s^2\alpha'$ is the ten-dimensional
Newton constant. $g_s$ and $\alpha'$ are the string coupling
constant and the inverse string tension respectively; $O(\alpha')$
represents the $\alpha'$ corrections (see the Appendix A).  From now
on we usually will suppress indexes of the tensor forms and wright
only their rank as a subscript; $F'_{(4)} = F_{(4)} + 2 A_{(1)}
\wedge H_{(3)}$.

   The action \eq{2a} correctly reproduces the low energy string amplitudes
for the massless modes. There are also various dual forms of the action
\eq{2a} expressed through the dual fields \eq{sd}. The latter fact is also
true for the Type IIB theory.

{\bf In the Type IIB theory:}\\
$\chi$ is the axion pseudo-scalar; $A_{\mu\nu}$ is the 2-form
potential; $A_{\mu\nu\alpha\beta}$ is the self-dual 4-form
potential. Their dual "magnetic" fields are:

\beqn
F_{\mu_1...\mu_9} = \epsilon_{\mu_1...\mu_{10}}
\diff_{\mu_{10}}\chi \nonumber\\
F_{\mu_1...\mu_7} = \epsilon_{\mu_1...\mu_{10}}
F^{\mu_8\mu_9\mu_{10}} \nonumber\\
F_{\mu_1...\mu_5} = \epsilon_{\mu_1...\mu_{10}}
F^{\mu_6...\mu_{10}}. \label{sd'}
\eeqn
The bosonic part of the low energy SUGRA action of the Type
IIB theory looks as follows \cite{GrScWi}:

\beqn
S_{IIB} = \frac{1}{16\pi\Gamma^{(10)}_N}\int d^{10}
x\sqrt{-G}\Bigl\{e^{-2\varphi}\left[\cR +
4\left(\nabla_{\mu}\varphi\right)^2 - \frac13 H^2_{(3)} \right] -
\nonumber\\ - \alpha'\left(\nabla_{\mu}\chi\right)^2 -
\frac{\alpha'}{3} F^2_{(3)} + ...\Bigr\} + O(\alpha') \label{2b}
\eeqn
Unfortunately, we do not known any canonical way of writing an action
for the self-dual field-strength $F_{(5)}$ of the potential
$A_{(4)}$.  This does not cause any problems because one knows
equations of motions in this Type IIB SUGRA theory
\cite{GrScWi}. That is enough for the consideration of classical
solutions.

   Now we continue with SUGRA solitons. These solitons correspond,
from the string world-sheet point of view, to some approximate
($\frac{\alpha'}{R^2} \to 0$ where $R$ is a characteristic size of
the solution) superconformal field theories.  We will concentrate
first on the solutions that preserve some SUSY, i.e.  on the charged
BPS solutions.  They are generalizations of the extreme
Reissner-Nordstrom solution to the case when curvature
singularities live on multi-dimensional world-volumes. The latter are
surrounded by multi-dimensional event horizons.

  These solutions will be extended {\it p-branes} of $p$ {\bf spatial}
dimensions. They can carry either "electric"  charge under
the $A_{(p+1)}$ form or "magnetic" one under the $A_{(7-p)}$ form. It is
these charges which are present in the ten-dimensional SUSY algebra \eq{susy}
as the central extensions with tensor indexes.

  A p-brane solution carries the charge under a $A_{n}$ or its dual
field if the p-brane world-sheet theory includes either $\int d^{n}
x \epsilon^{a_1...a_{n}} A_{\mu_1...\mu_{n}} \diff_{a_1}
x^{\mu_1}...\diff_{a_{n}} x^{\mu_{n}}$ or $\int d^{n} x
\epsilon^{a_1...a_{n}}\epsilon_{\mu_1...\mu_{10}}
A^{\mu_1...\mu_{10-n}} \\ \diff_{a_1} x^{\mu_{9-n}}...\diff_{a_{n}}
x^{\mu_{10}}$ terms, respectively. For example, a particle (0-brane)
carries the charge under an $A_{\mu}$ gauge potential if its
world-sheet theory includes the following term $\int d x_{\mu}
A_{\mu}$. A string (1-brane) carries the charge under a
$B_{\mu\nu}$ tensor potential if its world-sheet theory includes the
term:  $\int d^2\sigma \epsilon^{ab} B_{\mu\nu}\diff_a x^{\mu}
\diff_b x^{\nu}$.

   It is worth mentioning that no perturbative string state can be
charged with respect to any $R-\tilde{R}$ field. In fact, closed string vertex
operators contain only gauge invariant $F$'s rather than
corresponding potentials $A$ \cite{GrScWi} (see also the end of the
Appendix B).

\section*{4.1. Solitons in ten dimensions}

To get a static p-brane solution one starts with the ansatz for
the string metric \cite{DuKh94}:

\beq
ds^2 = h\left[f^{-1}_p\left(-dt^2 + dx^2_1 + ... +
dx^2_p\right) + dx^2_{p+1} + ... + dx^2_9\right]. \label{pb}
\eeq
It breaks the ten-dimensional Lorentz group $SO(9,1)$ down to the
$SO(p,1) \times SO(9-p)$. Hence, the $p$-brane should be dynamical
to maintain the invariance under the full group. Here $f_p$ and $h$
are some functions of the transverse coordinates $x_{p+1},...,x_9$
which should be found.  Also one allows the dilaton $\varphi$ and the
component $A_{0...p}$ of the corresponding $R-\tilde{R}$ field to be
non-zero, setting all other fields to zero.

   It is also possible to consider a 1-brane charged soliton, placed, for
example, in the 9-th direction and charged with respect to the
$NS-\tilde{NS}$ tensor field $B_{\mu\nu}$ \cite{DuKh94}.  This can be
done using \eq{pb} for the case of $p=1$ and keeping non-zero the
$B_{09}$ field instead of $A_{0...p}$. Such a soliton would be a massive
fundamental superstring excitation. It is in the perturbation spectrum
because its tension is of the order of one in the string
units.  One calls it the {\it F-string}, where F comes from the
"Fundamental".

  In a similar way one can construct an object which is
"magnetically" charged with respect to the field $B_{\mu\nu}$. This is a so
called {\it solitonic NS 5-brane} \cite{DuKh94}.  Which has the mass
per unit volume proportional to $\frac{1}{g^2_s}$
\cite{DuKh94}. Hence, the solitonic 5-brane is a non-perturbative excitation.

   In this paper we mostly will be interested in the $R-\tilde{R}$
solitons, i.e. charged with respect to the $R-\tilde{R}$ tensor
fields.  Because they, unlike the F-string and NS 5-brane,
have an exact conformal field theory treatment (see below). We mean that one
knows what kind of objects in string theory quantize these solutions.

   The main distinction of the $R-\tilde{R}$ solitons from the
F-string and NS 5-brane comes from the fact that their mass per unit
volume in the string units is of the order of $\frac{1}{g_s}$ \cite{DuKh94}.
This is due to the unusual power of the exponential of the dilaton in front
of the kinetic terms \eq{2a},\eq{2b} for the $R-\tilde{R}$ potentials
\cite{GrScWi}.

    The p-brane $R-\tilde{R}$ solution (metric \eq{pb},
$\varphi$ and $A_{0...p}$ fields) becomes BPS if it preserves some
part of the SUSY transformations in the corresponding SUGRA theory
\cite{DuKh94}.  It means that the part of the SUSY transformations
does not move the fields under consideration. Hence, to obey this
condition one should insist on the vanishing of the SUSY transformations for
the {\it gravitino} and {\it dilatino}\footnote{Superpartners of the graviton
and dilaton respectively.} (which were taken to be zero) in the
corresponding Type II SUGRA theory \cite{DuKh94}.  This forces
the dilaton, $R-\tilde{R}$ tensor field and metric to be related
to each other and to take the form:

\beqn
ds^2 = f^{-\frac12}_p\left(-dt^2 + dx^2_1 + ... + dx^2_p\right) +
f^{\frac12}_p \left(dx^2_{p+2} + ... + dx^2_9\right), \nonumber\\
e^{-2\varphi} = f_p^{\frac{p-3}{2}}, \nonumber\\
A_{0...p} = - \frac12 \left(f^{-1}_p - 1 \right),
\label{sol}
\eeqn
where $p=0,2,4,6,8$ in the Type IIA and $p=1,3,5,7$ in the Type IIB
theories \cite{DuKh94} (see the Appendix B and previous
subsection).

  All these solutions are BPS and for any function $f_p$ they
preserve only half of the SUSY because of the conditions:

\beq
\eta_r = \gamma^0...\gamma^p \eta_l
\eeq
where $\eta_r$ and $\eta_l$ are the right and left chiral
parts of the infinitesimal local SUSY generator. The equations
of motion of the theory (related to the closure of the SUSY algebra)
imply \cite{DuKh94} that $f_p$ should be a harmonic
function $\Delta^{(p)} f_p = 0$. Here $\Delta^{(p)}$ is the flat
Laplasian in the directions $p+1,...,9$.  From \eq{sol} one gets the
extreme limit of the charged {\it black} p-brane when the harmonic
function is

\beq
f_p = 1 + \frac{n c^{10}_p}{r^{7-p}}, \label{f}
\eeq
where $r = \sqrt{x^2_{p+1} + ...  +
x^2_9}$; $c^{10}_p$ is related to the minimum charge of the
p-brane and will be accounted for particular solutions in the section
five.  The corresponding charge is defined through a kind of the
integral Maxwell equation for a tensor gauge potential:

\beq
Q_p \sim \int e^{2\varphi} * F_{(p+2)}
\eeq
where $F$ is the field strength of the $A$; "$*$" is the {\it
Hodge} dual operation which is done through the absolutely
anti-symmetric tensor in ten dimensions. The integral is taken over a
hupersurface which surrounds the p-brane.

  In the eq. \eq{f} $n$ is an integer because of the Dirac type
quantization of charges \cite{DuKh94}:

\beq
Q_p Q_{6-p} = 2\pi k, \quad k\in {\bf Z}.
\eeq
where $Q_{6-p}$ is the charge of the dual brane soliton with respect
to the dual field.

   Since all BPS solutions treated so far depend on some harmonic
function $f_p$, one can construct multiple brane solutions. It is done by
taking $f_p = 1 + \sum_i \frac{c_p}{\left|{\vec r} - {\vec r}_i\right|^{7-p}}$
which describes a set of branes at positions ${\vec r}_i$
\cite{DuKh94}.  Such solutions similarly break
only half of the SUSY. Moreover, they are in the static equilibrium
because the force between {\bf parallel BPS solutions with the same
parity of charges} is always zero \cite{DuKh94}. In fact, the gravitational
attraction between the BPS solitons is compensated by the repulsion
through the $R-\tilde{R}$ tensor fields.

    One also can construct BPS bound states of $p$-branes with
different $p$'s obeying some conditions \cite{Du}. As we explain in the
subsection 4.3, such a construction already breaks more than half of
the SUSY. The conditions on $p$'s are needed to preserve at
least some smaller part of it. We use this construction below when
consider multicharged black holes.

\section*{4.2. Black Holes from Black Branes}

   Now let us proceed with the Type II theory compactified to $d$
dimensions on a torus $T^{10-d}$ \cite{GrScWi}. We identify
coordinates by

\beq
x_i \sim x_i + 2\pi R, \quad i = d,...,9, \label{id}
\eeq
and choose the periodic boundary conditions on this $10-d$
dimensional "box".  Fields that vary over the box will acquire masses
of the order of $\frac{1}{R}$ where $R$ is the typical
compactification size\footnote{Which usually is taken to be big in
comparison with $\sqrt{\alpha'}$ to maintain the $\alpha'$
corrections to be small (see the Appendix A).}. Thus, if we are
interested in the {\bf low energy} physics in $d$ extended
dimensions, fields might be taken to be independent of the internal
coordinates of the torus.

    One can observe that if we have any solution in ten dimensions
which is periodic under $x_i\to x_i + 2\pi R$, then it will also be a
solution of the compactified theory. For any $p$-brane, the solution
is automatically translation invariant in the directions parallel to
the brane.

  We will be interested in solutions where a brane is
completely wrapped along the internal directions. Its volume is
represented as a part of the torus $T^{10-d}$.  Therefore, from the
point of view of an observer in $d$ dimensions one has a
localized, spherically symmetric solution.  The latter can
carry a charge with respect to a gauge field which is the remnant

\beq
A_{\mu} = A_{\mu i_1...i_p} \label{abel}
\eeq
of the corresponding $R-\tilde{R}$ tensor field. Here $\mu$ is the $d$
dimensional index and $i_n$ lie in compact directions.

   The BPS solutions of the kind under consideration correspond to the extreme
limits of the charged black holes.  The final result is that the
$d$-dimensional solutions are given again by \eq{sol} with $p=0$ but
now in terms of $d$-dimensional harmonic functions. Hence, when we
are in the $d$-dimensional theory, the only way we have to tell that
a black hole contains a particular type of the $p$-branes is by
looking at the gauge fields that it excites.

\section*{4.3. $R-\tilde{R}$ solitons as $D$-branes}

     In this section we show what kind of objects in superstring
theory quantize the $R-\tilde{R}$ black $p$-branes and, hence, the
black holes with corresponding kind of charges.

   Originally it was thought that the only allowed open
string theory is the non-orientable Type I theory in ten dimensions
with the gauge group $SO(32)$ \cite{GrScWi}. This open string theory
has the {\it Neumann (N)} boundary conditions on all coordinates:

\beq
n^a \diff_a x_{\mu} = 0 , \quad \psi^{\mu} = \pm \tilde{\psi}^{\mu},
\qquad \mu = 0,...,9
\eeq
where $n^a$ is a vector normal to the boundary; $\psi^{\mu}$ are
the world-sheet superpartners of the coordinates $x_{\mu}$. The latter
represent positions of the string in the target space.

    However, one actually can make open string sectors in the
closed Type II superstring theory \cite{Pol}.  As it
appears, the open strings in these sectors should also have
the {\it Dirichlet (D)} boundary conditions on some part of the
string coordinates:

\beqn
n^a \diff_a x_m = 0 , \quad \psi^m = \pm \tilde{\psi}^m,
\quad m = 0,...,p \quad (N)\nonumber\\
x^i = C^i, \quad \psi^i = \mp \tilde{\psi}^i \quad i = p+1,...,9 \quad
(D), \label{bou}
\eeqn
where $C^i$ are arbitrary, fixed numbers. Therefore, there are
open strings with their boundaries allowed to lie only on
$p+1$-dimensional sub-manifolds of the ten-dimensional target space.
While in the bulk of this space one has only ordinary Type
II closed strings. These sub-manifolds are referred to as {\it
Dp-branes}. Here $p=0,2,4,6,8$ for the Type IIA and
$p=1,3,5,7$ for the Type IIB theories \cite{Pol}.

   The Dp-branes have several features which are interesting for
us.  First, they break the Lorentz invariance in the target space.
Hence, to maintain it, one should consider these Dp-branes as
dynamical excitations.  Second, because of the boundary conditions
\eq{bou}, the Dp-branes break a half of the SUSY
transformations\footnote{As in the case of the Type I superstring theory,
the SUSY generators are related to each other on the boundary \cite{GrScWi}}.
Thus, the Dp-branes are BPS states.  Third, the closed Type II strings are
charged with respect to the $B_{\mu\nu}$ field (see the Appendix A) so that
the two-dimensional theory is invariant under the transformations:

\beq
B_{\mu\nu} \to B_{\mu\nu} + \diff_{[\mu} \rho_{\nu]}. \label{bb}
\eeq
When we break the string, however, there is a boundary term
appearing after such a transformation. Hence, to maintain the
invariance of the theory one has to accept that the boundary is
charged with respect to some abelian gauge field $A_{m}$. In this case
the boundary term under consideration can be compensated by a shift
$A_{m}\to A_{m} - \rho_{m}$. (This shift is different from the
ordinary gauge transformation $A_{m} + \diff_{m}\lambda$ of the field
$A_{m}$.)  Therefore, there should be the gauge fields living on the
Dp-branes.  Below we give other arguments in favor of their
appearance.

     The presence of a Dp-brane in the Type II superstring theory can be
described in particular gauge by the superconformal field theory \cite{Lei89}
with additional boundary terms:

\beqn
S_{2d} = \frac{1}{2\pi \alpha'}\int d^2\sigma
\left(G^{\mu\nu}\diff_a x_{\mu}\diff^a x_{\nu} +
\epsilon^{ab}B^{\mu\nu}\diff_a x_{\mu}\diff^b x_{\nu}\right)
+ S_{boundary} + ...,
\nonumber\\ S_{boundary} = \int ds \left[\sum^p_{m=0}
A_m(x_0,...,x_p)\diff_t x^m +
\sum^9_{i=p+1}\phi_i(x_0,...,x_p)\diff_n x^i\right], \label{sup}
\eeqn
where dots in the first equation stand for the fermionic
superpartners; $s$ is some parameter on the boundary; $\diff_t$ and
$\diff_n$ are the tangential and normal derivatives to the boundary
respectively; $A_{m}$ is the above mentioned gauge field on the
world-volume of the Dp-brane and $\phi_i$ are coordinates which
determine positions of the Dp-brane in the bulk.

  The low energy dynamics of the Dp-branes is governed by low energy {\it
Born-Infeld} action \cite{Lei89}:

\beqn
S_{BI} = - M_p \int d^{p+1}\xi
e^{-\varphi}\left[\sqrt{\det\left(g_{mn} + b_{mn} + 2\pi
F_{mn}\right)} - 1 \right] + ... \nonumber\\ F_{mn} = \diff_{[m} A_{n]},
\quad g_{mn} = G_{ij}\diff_m \phi_i\diff_n \phi_j + G_{im}\diff_n\phi_j +
G_{mn}, \nonumber \\ b_{mn} = B_{ij}\diff_m\phi_i\diff_n\phi_j +
B_{i[m}\diff_{n]}\phi_j + B_{mn} \label{BI}
\eeqn
which correctly reproduces
the low energy string amplitudes for the massless modes. Here again we
skipped fermionic terms. One can calculate the mass $M_p$ per unit volume,
observing that there is a coupling of the Dp-brane to the metric from the
bulk space.  This coupling is another evidence in favor of the fact that
Dp-branes are dynamical excitations.  One finds that \cite{Pol} $M_p =
\frac{\pi}{g_s} \left(4\pi^2\alpha'\right)^{\frac{3-p}{2}}$, which is the
first indication that the Dp-branes are $R-\tilde{R}$ solitons.  In fact,
among all the string excitations only these solitons have tensions
proportional to $\frac{1}{g_s}$.  Actually, one can prove \cite{Pol95} that
the Dp-branes do carry charges with respect to the $R-\tilde{R}$ tensor gauge
fields.  In fact, the force between any two equivalent Dp-branes vanishes
\cite{Pol}, because they are BPS solitons. This is an indication that there
should be a repulsion, due to some tensor fields, compensating the
gravitational attraction.

   What physics underlies the connection between "flat" Dp-branes and
"curved" $R-\tilde{R}$ $p$-brane solitons?  The $R-\tilde{R}$ black
$p$-branes which were considered in the previous section are
solutions of the SUGRA equations of motion as $\frac{\alpha'}{R^2}\to
0$. Here $R$ is a characteristic size of the solution.  At the same time the
Dp-branes can be exactly described as the manifolds on which strings
can terminate. The description is valid when $g_s \to 0$. In the latter limit
the size of the event horizon becomes smaller than the string characteristic
scale $\sqrt{\alpha'}$. Thus, the Dp-branes give a microscopic, string
theory, description of the $R-\tilde{R}$ solitons. It is believed that when
$g_s$ is not zero, the Dp-branes are dressed and, by curving
space-time, form an event horizon.

    What about the low energy theory for the Dp-branes? At low
energies the fields in the action \eq{BI} are week and in the flat
target-space we have:

\beq
S_{BI} = \frac{1}{4} F^2_{mn} + |\diff_m\phi^i|^2 + ...\label{bibi}
\eeq
where dots stand for the fermionic terms. Therefore, the
theory for the low energy excitations of a Dp-brane is described by
the compactification \cite{Wit95} of the ten-dimensional SUSY QED to
$(p+1)$ dimensions. In fact, in diverse dimensions the scalars $\phi_i$ can
be treated as the components $A_i$ of the ten-dimensional vector field
$A_{\mu}$ along compact directions.

   The low energy action \eq{bibi} also can be found from
another point of view \cite{Wit95}. In fact, at low energies
(or as $\alpha' \to 0$) the string which terminates on the brane
looks as the massless vector excitation - the lowest energy
excitation of the open string \cite{GrScWi}. At this point one easily
finds that the low energy theory for such an excitation is SUSY QED -- the
only supersymmetric and gauge-invariant action containing smallest number of
the derivatives.

   The last point of view helps to understand the low energy theory
describing a bound state of the Dp-branes. Let us consider $n$ parallel
Dp-branes with the same $p$. In this case, in addition to the strings which
terminate by both their ends on the same Dp-brane, there are strings
stretched between different ones. The strings of the first kind give
familiar massless vector excitation on each Dp-brane.  While the strings of
the second kind give massive (mass proportional to a distance
between corresponding Dp-branes) vectors. They are charged with respect to
the gauge fields on the both Dp-branes. Therefore, these latter vector
excitations are similar to the $\cW^{\pm}$-bosons.  They acquire masses
through a kind of the Higgs mechanism (splitting of Dp-branes) and become
massless when Dp-branes approach each other.  Hence, one finds that the
world-volume theory on the bound state of $n$ Dp-branes is nothing but $U(n)
= SU(n) \times U(1)$ maximally supersymmetric {\it SUSY Yang-Mills (SYM)}
theory \cite{Wit95}.  While all possible positions $\phi^i$ of the Dp-branes
are the Higgs vacuum expectation values. They parametrize flat directions of
the corresponding scalar potential in the SYM and the $U(1)$ factor
corresponds to the center of mass position of the bound state.

    In this language one also can understand how to construct bound
states of the D-branes with different $p$'s \cite{Du}. For example, SYM
theory on a Dp-brane has BPS excitations which break half of the SUSY
on the world-volume of the Dp-brane.  In four dimensions ($p=3$)
these are instantons. While when one has $p\ge 3$ the instantons
acquire extra (more than 0) longitudinal dimensions. They
become monopoles at $p=4$, strings at $p=5$ and etc..
One interprets the Dp-branes with such BPS excitations as Dp-brane
bound states. The latter break more SUSY than "mother" brane and have
given above relation between their values of $p$'s.  One also can use
other constructions, with other relations between $p$'s, which we do
not need in this review.

   Also it is possible to understand, in analogy with the case of
strings, what kinds of branes can terminate on each other
\cite{StTo95}.  In fact, the strings terminate on the Dp-branes
because their boundaries can be considered as particles living on the
Dp-brane.  These particles carry electric charges with respect to
the gauge fields which also live on the Dp-branes.  Similarly, if
there is a tensor field $A_{m_1...m_n}$ living on a Dp-brane then a
D(n+1)-brane can terminate on it. For example, in the Type IIA theory
the solitonic NS 5-brane contains the $B_{mn}$ field in its
world-volume theory.  Therefore, a D2-brane can terminate on this
5-brane. In this case the D2-brane boundary is a string living on the NS
5-brane and charged with respect to the $B_{mn}$ field.  We use such
a construction when consider four-dimensional black holes in the
section five.

\section*{5. Microscopic black hole entropy}

  In this section we consider a concrete extreme black hole
solution in a compactification of the Type II superstring theory.
We calculate the semiclassical value of the area of its horizon.
After that we identify what Dp-brane bound state quantizes this black
hole.  Then we count the degeneracy of the state and find a complete
equivalence between log of this number and the horizon area.

    We need to work with the extreme BPS solutions which have a
singularity of any field defining the solution shifted from the event
horizon. If this is not the case, then the $\alpha'$ corrections to
the black hole area are strong. As it appears \cite{Mal96}, to
have a non-singular event horizon, one needs less parameters for
five-dimensional black holes than for four-dimensional ones.
Therefore, for simplicity, we start with a five-dimensional
extreme black hole.  Then we follow with the calculation of the
entropy and decay rate of a five-dimensional non-extreme black
hole. After that we sketch similar calculations for a four-dimensional
extreme black hole.

\section*{5.1. The five-dimensional black hole}

  In the following three subsections we closely follow the
presentation in the ref.  \cite{Mal96}.  We consider the Type IIB
string theory compactified on a torus $T^5$.  The low energy theory
is the maximally supersymmetric SUGRA theory in five dimensions: it has 32
components of the SUSY generators.  The theory contains 27 abelian gauge
fields, appearing as in \eq{abel} from the metric and various antisymmetric
tensor fields.  The full string theory contains charged objects that
couple to each combination of these fields. Due to the Dirac condition
these charges should be quantized in integer multiples of the elementary
units.

   Thus, we consider the black hole solution which, in the extreme
limit, is a bound state of the $Q_5$ $R-\tilde{R}$ 5-branes wrapped
on the $T^5$ and of the $Q_1$ $R-\tilde{R}$ 1-branes wrapped on a
$S^1$ (we choose it as the direction 9) carrying quantized
momentum\footnote{$N$ is the charge with respect to the gauge field
$G_{\mu 9}$.} $P=\frac{N}{R_9}$ along the compact direction of
the $R-\tilde{R}$ 1-brane.  Such a bound state of the $R-\tilde{R}$
p-branes should preserve a half of the SUSY transformations for each
kind of the gauge charges. Hence, this BPS solution preserves
$\frac18$ of all $32$ components of the SUSY generators.

    We start by presenting the non-extreme ten-dimensional solution
\cite{HoMaSt96}:

\beqn
e^{-2(\varphi - \varphi_0)} = \left(1 + \frac{r^2_0 sh^2(\gamma)}{r^2}
\right) \left(1 + \frac{r^2_0 sh^2(\alpha)}{r^2}\right)^{-1}
\label{dila}
\eeqn
where $\varphi_0$ is the vev of the dilaton $\varphi$ and

\beqn
ds^2_{str} = \left(1 + \frac{r^2_0
sh^2(\gamma)}{r^2}\right)^{-\frac12} \left(1 + \frac{r^2_0
sh^2(\alpha)}{r^2}\right)^{-\frac12} \times \nonumber\\
\times \Bigl[-dt^2 + dx_9^2 + \frac{r_0^2}{r^2}\left(ch(\sigma) dt +
sh(\sigma) dx_9\right)^2 + \nonumber\\ + \left(1 + \frac{r^2_0
sh^2(\alpha)}{r^2}\right) \left(dx_5^2 + ... + dx_8^2\right)\Bigr] +
\nonumber\\ + \left(1 + \frac{r^2_0
sh^2(\gamma)}{r^2}\right)^{\frac12} \left(1 + \frac{r^2_0
sh^2(\alpha)}{r^2}\right)^{\frac12}\left[\left(1 - \frac{r_0^2}{r^2}
\right)^{-1} dr^2 + r^2 d\Omega_3^2\right].  \label{baho}
\eeqn
Here the subscript $"str"$ means that the line element is written in
the string frame (see the discussion below the eq. \eq{26} in the
Appendix A); $d\Omega^2_3$ represents the angular part of the metric
for the polar angles of the vector ${\vec r}$, where $r = \sqrt{x_1^2
+ ... + x_4^2}$. Also components of the $F_{\mu\nu\rho}$ are non-zero
since the solution carries the 1- and 5-brane charges. The last
charge is dual to the first one due to the relation \eq{sd'}.

   This solution parametrized by four independent quantities:
$\alpha,\gamma,\sigma,r_0$. There are also two extra parameters which enter
through the charge quantization conditions.  These parameters are the radius
of the 9-th dimension $R_9$ and the product of the radii in the other four
compact directions $V = R_5 R_6 R_7 R_8$.  The three charges of the black
brane can be most easily viewed when the ten-dimensional solution \eq{baho}
compactified to six dimensions:

\beqn
Q_1 = \frac{V}{4\pi^2 g_s} \int e^{2\varphi_6} * F_{(3)} =
\frac{Vr_0^2}{2g_s} sh(2\alpha) \nonumber\\
Q_5 = \frac{1}{4\pi^2 g_s} \int F_{(3)} = \frac{r_0^2}{2g_s}
sh(2\gamma) \nonumber\\ N = \frac{R_9^2Vr_0^2}{2g^2_s}
sh(2\sigma),\label{char}
\eeqn
where "$*$" is the {\it Hodge} duality operation in six
dimensions $t,...,x_5$. The integrals are taken over a
three-dimensional sub-manifold surrounding the black hole;
$\varphi_6$ is the six-dimensional dilaton. It differs from the
ten-dimensional one by a linear combination of logs of components of
the ten-dimensional metric along compact directions.  For simplicity
we set from now on $\alpha' = 1$.  All the charges are normalized to
be integers.

   Now reducing \eq{baho} to five dimensions, using the
standard dimensional reduction procedure \cite{MaSc93}, the solution
takes simple and symmetric form:

\beq
ds_5^2 = - \Delta^{-\frac23}h dt^2 + \Delta^{\frac13}\left(h^{-1}
dr^2 + r^2d\Omega^2_3\right) \label{five}
\eeq
where

\beqn
\Delta = \left(1 + \frac{r_1^2}{r^2}\right)  \left(1 +
\frac{r_5^2}{r^2}\right) \left(1 + \frac{r_n^2}{r^2}\right)
\nonumber\\
h = 1 - \frac{r_0^2}{r^2} \nonumber\\
r_1^2 = r_0^2 sh^2(\alpha), \quad r_5^2 = r_0^2 sh^2(\gamma), \quad
r_n^2 = r_0^2 sh^2(\sigma). \label{r_n}
\eeqn
This is just the five-dimensional Schwarzschild metric with the time
and space components rescaled by different powers of the function
$\Delta$.  The event horizon is at the surface $r=r_0$. Note that the
five-dimensional Reissner-Nordstrom solution corresponds to the case
of $\alpha=\gamma=\sigma$.

   Why did we chose this solution? The reason is that three charges, in the
case under consideration, is the minimal number needed to have a non-singular
solution at the event horizon \cite{Mal96}.  In fact, as we have seen
in the section 4.1, a black p-brane produces the dilaton field of the
form $e^{-2\varphi} = f_p^{\frac{p-3}{2}}$, with $f_p$ being a
harmonic function.  A superposition of the black branes produces the
product of such functions and one sees how 1-branes can cancel
5-branes in their effect on the dilaton \eq{dila}.  A similar thing
is true for the compactification volume.  For any p-brane, the string
metric is such that as we get closer to the brane the volume parallel
to it shrinks, due to the brane tension. While the volume perpendicular
to it expands, due to the pressure of the electric field lines.  It
is easy to see how superposing the 1- and 5-branes can stabilize the
volume in the directions 5,6,7,8, since they are perpendicular to the 1-brane
and parallel to the 5-brane.  The volume in the direction 9 would still seem
to shrink, due to the tension of the branes. This is indeed why we put the
momentum along the 1-branes, to balance the tension and produce a stable
radius in the 9-th direction.

  Also there are {\it U-duality} \cite{WiHuTo95} transformations
which exchange (see the Appendix C) the above mentioned 27 gauge
fields with each other and invert different couplings (such as $g_s,
R_9$ and $V$) in the theory.  Therefore, via these U-duality transformations
one can get from our solution another one which is charged with
respect to any three of that 27 charges.

  One can calculate thermodynamic quantities corresponding to
this solution \cite{Mal96}. For example, the mass is equal to

\beq
M = \frac{R_9Vr_0^2}{2g_s^2} \Bigl(ch(2\alpha) + ch(2\gamma) +
ch(2\sigma)\Bigr). \label{admm}
\eeq
While the entropy is:

\beq
S = \frac{A_{10}}{4\Gamma^{(10)}_N} = \frac{A_5}{4\Gamma^{(5)}_N} =
\frac{2\pi R_9Vr_0^3}{g_s^2} ch(\alpha) ch(\gamma) ch(\sigma)
\label{ent'}
\eeq
where the five-dimensional Newton constant is $\Gamma^{(5)}_N
=\frac{\pi g_s^2}{4VR_9}$. And the temperature of this non-extreme
black hole is equal to:

\beq
T = \frac{1}{2\pi r_0 ch(\alpha) ch(\gamma) ch(\sigma)} \label{temper}
\eeq
Let us take a look at the formulae \eq{admm}, \eq{ent'} and
\eq{temper}. It is possible to trade the six parameters of the
general solution  for the six quantities
$(N_1,N_{\bar{1}},N_5,N_{\bar{5}}, N_l, N_r)$ which are the
"numbers" of the 1-branes, anti-1-branes, 5-branes, anti-5-branes,
left-moving momentum and right-moving momentum respectively. This is
accomplished by equating the total mass, charges and the entropy of
the black hole with those of a collection of
the numbers $(N_1,N_{\bar{1}},N_5,N_{\bar{5}}, N_l, N_r)$
non-interacting "constituent" branes, anti-branes and momentum. By
non-interacting we mean that the mass is simply the sum of the masses
of the constituents.  We take the $N$'s to be

\beqn
N_1 = \frac{Vr_0^2}{4g_s} e^{2\alpha}, \qquad N_{\bar{1}} =
\frac{Vr_0^2}{4g_s} e^{-2\alpha}; \nonumber\\
N_5 = \frac{r_0^2}{4g_s} e^{2\gamma}, \qquad
N_{\bar{5}} = \frac{r_0^2}{4g_s} e^{-2\gamma}; \nonumber\\
N_r = \frac{r_0^2R_9^2V}{4g_s^2} e^{2\sigma}, \qquad
N_l = \frac{r_0^2R_9^2V}{4g_s^2} e^{-2\sigma}.\label{cond}
\eeqn
These $N$'s reduce to the numbers of branes, anti-branes and
momentum in certain limits where those concepts are well defined
\cite{Mal96}.  In terms of them the charges are simply $Q_1 = N_1 -
N_{\bar{1}}, Q_5 = N_5 - N_{\bar{5}}, N = N_r - N_l$, the total
energy is:

\beq
M = \frac{R_9}{g_s}(N_1 + N_{\bar{1}}) + \frac{R_9V}{g_s}(N_5 +
N_{\bar{5}}) + \frac{1}{R_9}(N_r + N_l), \label{noext}
\eeq
and the volume and radius are

\beqn
V = \left(\frac{N_1N_{\bar{1}}}{N_5N_{\bar{5}}} \right)^{\frac12},
\quad R_9 =  \left(\frac{g_s^2N_rN_l}{N_1N_{\bar{1}}}\right)^{\frac14}.
\eeqn

    Of course there seems to be no reason for neglecting interactions
between collections of branes and momentum modes composing a highly
non-extreme black hole at strong or intermediate coupling. Hence,
the conditions \eq{cond} would seem to be inappropriate for
describing a generic black hole. However, the utility of these
definitions can be seen when we reexpress the black hole entropy
\eq{ent'} in terms of the $N$'s. It takes the remarkably simple form

\beq
S = 2\pi \left(\sqrt{N_1} + \sqrt{N_{\bar{1}}}\right)
\left(\sqrt{N_5} + \sqrt{N_{\bar{5}}}\right)
\left(\sqrt{N_l} + \sqrt{N_r}\right) \label{noext'}
\eeq
and appears, being U-duality invariant, to be useful and transparent
for the understanding of physics below.

   The extreme limit corresponds to taking $r_0\to 0$ and
$\alpha,\gamma,\sigma \to \infty$, keeping the charges \eq{char}
finite. Thus, in terms of \eq{cond}, we include either branes or
anti-branes rather than both of them. The extreme solution preserves
only $\frac18$ of the SUSY transformations and corresponds to the
short supermultiplet.  In the extreme limit the entropy becomes:

\beq
S = 2\pi\sqrt{NQ_1Q_5} \label{ent}
\eeq
and the temperature vanishes.

   Note that the entropy of the BPS solution is independent, as we
discussed in the section three, of any continuous parameters.
Therefore, it is independent of $\varphi_0$ and one says that the
curved black hole description is valid at $g_s =
e^{\varphi_0}\sim 1$ (see the Appendix A), otherwise (if $g_s\to 0$
or $g_s\to\infty$) the string perturbation or flat Dp-brane
description would be applicable\footnote{If $g_s\to\infty$, then
perturbation theory of the dual string would be applicable, because
as we sketched in the Appendix C all the known superstring
theories are related to each other through a strong-week ($g_s\to
\frac{1}{g_s}$) or other kinds of duality.}.  At the same time in the
black hole region all sizes should be big in comparison with
the string scale to make the $\alpha'$ corrections small!  Thus, in
the above calculation, using the low energy theory of the Type
IIB string, we adopted the following approximation ($\alpha'=1$):

\beq
g_sQ_1>>1, \quad g_sQ_5>>1, \quad g^2_sN>>1. \label{aprox}
\eeq
Here $g_sQ_1, g_sQ_5$ and $g^2_s N$, as one can see from the metric
\eq{baho}, set characteristic size of the black hole, i.e. the area
of event horizon and its radii.

   It is worth mentioning that the formula \eq{ent} for the entropy
is U-duality invariant: after any U-duality transformation the value
of the charges will remain the same, but we would have a bound state
of different kinds of branes. This U-duality invariance and
independence upon $\varphi_0$ is the indication of the fact that one
is able to find a complete agreement between calculations of the
entropy at different values of the couplings in the theory.

\section*{5.2. D-branes and the extreme black hole}

    We continue with the Type IIB string theory on the torus $T^5 =
T^4 \times S^1$. We consider a configuration of the $Q_5$ D5-branes
wrapping the whole $T^5$, $Q_1$ D-strings wrapping the $S^1$ and
momentum $\frac{N}{R_9}$ along 9-th direction. All charges are
integers. This configuration of the Dp-branes corresponds (by
comparison of the charges and of the mass) to the above described
bound state of the $R-\tilde{R}$ solitons.

   Now we use open string perturbation theory, i.e. we consider another
vacuum background for the Type IIB perturbation theory.  Then one should take
the effective coupling constants\footnote{$Q_1, Q_5$ and $N$ appear in the
effective couplings of the open string perturbation theory as
the multiplicities of the ends of the strings.} $g_sQ_1, g_sQ_5$ and
$g_s^2N$ to be much smaller than one!  This should be
contrasted with the approximations \eq{aprox} adopted in the previous
subsection.

   The total mass of the system is

\beq
M = \frac{Q_5 R_9 V}{g_s} + \frac{Q_1 R_9}{g_s} + \frac{N}{R_9}
\eeq
and saturates the corresponding BPS bound. We will calculate the
degeneracy of this state. Such a calculation was first done in the
ref. \cite{StVa96}.

   As the  Dp-branes are invariant under the Lorentz
transformations along the directions parallel to their volume,
they can not carry the momentum $\frac{N}{R_9}$ just moving rigidly.
Our task would be to identify the excitations which carry this
momentum. The BPS mass formula for the whole system implies that
these excitations have to be massless and moving along the $S^1$. In fact,
the excitation energy, defined as the total mass of the system
minus the mass of the 1-branes and 5-branes, is equal to the
momentum. If any excitation fails to be massless it would contribute
more to the energy than to the momentum and the BPS mass formula
would be violated.

   Excitations of the branes are described by massless open strings.
There are many types of the open strings to consider: those that go
from one 1-brane to another 1-brane, which we denote as $(1,1)$
strings, as well as the corresponding $(5,5)$, $(1,5)$ and $(5,1)$
strings (the last two being different because the strings are
oriented). We want to excite these strings and make them carry
the momentum in the direction of the $S^1$. However, exciting some of
them makes others massive \cite{Mal96}. Therefore, we
have to find how to excite the strings so that maximum number of
them remains massless. This configuration will have the highest
entropy.

     We have already said that the low energy Lagrangians for the
$(1,1)$ and $(5,5)$ strings are dimensional reductions of the $d=10$,
$N=1$ SYM to two or six dimensions for the D1- or D5-branes,
respectively.  However, because of the additional braking of the SUSY
due to the composition of different types of the Dp-branes the situation is
changed. One has the $N=4$ rather than $N=8$ SYM theory on the world-volume
of the D1-brane and $N=1$ rather than $N=2$ SYM on the D5-brane.

   The $(1,1)$ strings represent vector multiplets containing
$A^{(1)}_m$ gauge field with four scalars $\phi^{(1)}_i$ and
hypermultiplets in the adjoint representation containing four scalars
$\varphi^{(1)}_j$.  These scalars represent positions of the
D1-branes in the bulk.  At the same time the $(5,5)$ strings form the
vector multiplets containing only $A^{(5)}_n$ gauge field and the
hypermultiplets in the adjoint representation containing four scalars
$\varphi^{(5)}_j$. The latter represent positions of the D5-branes in the
bulk.

     One also can show (see, for example, \cite{Mal96}) that the
$(1,5)$ and $(5,1)$ strings form together the hypermultiplets.
They transform as the products of the fundamental
representation of the $U(Q_1)$ and the anti-fundamental of the
$U(Q_5)$ and their complex conjugate. These hypermultiplets contain
four scalars $\chi_i$ which represent relative positions of the D1-
and D5-branes.

    Thus, knowing the multiplet content, one can easily derive
low energy actions on the D1- and D5-branes with the mentioned numbers of
SUSY. The interaction Lagrangian for the mentioned fields is fixed largely by
the SUSY.  The only allowed coupling between the vector and hypermultiplets
is the gauge one.  The SUSY requires, however, some potential for the
scalars.  It arises, in our case, from the {\bf three} $D$-fields \cite{WeBa}
(do not mix with D-branes) for each gauge generator $t^a$.

   Let us count the number of massless bosonic excitations, the
number of fermionic excitations being equal to the bosonic ones due
to the SUSY.  The BPS states under consideration have only
left moving excitations. Classically one can view these states as
traveling waves propagating along the $S^1$. In order to have the
traveling wave solutions mass terms have to vanish
exactly.  If we set all fields in the Lagrangian to zero then we can
have the traveling waves for any field.  However, if we have a wave
for one field the mentioned potential terms generate effective mass
terms for other waves.

   If we give some expectation value to the scalars from the
$(1,5)$ and $(5,1)$ strings, then we are effectively separating the
D1- and D5-branes and we expect a small number of the massless
particles (proportional to $Q_1 + Q_5$). On the contrary a
configuration with the large number of the massless particles is
achieved by exciting all hypermultiplets, from
the $(1,1)$, $(5,5)$ and $(1,5)$, $(5,1)$ strings.  This gives masses
to the scalars from the vector multiplets describing the
transverse motion of the branes.  The total number of components of the
hypermultiplets is $4Q^2_1 + 4Q^2_5 + 4Q_1Q_5$.  Conditions of
vanishing of the scalar potential impose $3Q^2_1 + 3Q^2_5$
constraints. In addition we should identify gauge equivalent
configurations. The number of possible gauge transformations is
$Q^2_1 + Q^2_5$. This implies that the remaining number of bosonic
massless degrees of freedom is $4Q_1Q_5$. The counting, as we have
done, is correct for the large charges up to possible subleading
corrections.

  In order to calculate the entropy we notice that we have the gas of
the left moving particles with $N_{F,B} = 4Q_1Q_5$ bosonic and
fermionic species on the compact one-dimensional space of the length
$L=2\pi R_9$. Its energy is $E = \frac{N}{R_9}$. In the string theory the
degeneracy of such a state (as we sketched in the Introduction) is
\cite{Car86}:  $n \sim \exp{\left(\sqrt{\pi(2N_B + N_F)
\frac{EL}{6}}\right)}$ where under the square root in the exponent there is
the number of oscillator states. Then the entropy of the state is equal to:

\beq
S = \log{n} = 2\pi \sqrt{Q_1Q_5N}
\eeq
which is in the perfect agreement, including numerical coefficient,
with \eq{ent}.

\section*{5.3. D-branes and the non-extreme black hole}

  In this subsection we discuss the non-extreme black hole in the
D-brane picture \cite{HoMaSt96,MaSt96,HoSt96}.  We are working in the
{\it dilute gas} regime

\beq
r_0, r_n << r_1, r_5, \label{apr}
\eeq
where the $r$'s are defined in the eq. \eq{r_n}. Under these
conditions one is very close to a configuration of the extreme D1-
and D5-branes and SUSY non-renormalization arguments do indeed help
us \cite{MaDa96} (see also the second paper in \cite{Str96}) and explain the
agreement that we are going to find.  This is the simplest case of the near
extreme black hole, if we want to consider more general ones, one
has to include other excitations besides the right movers.

   The mass of the state is equal to:

\beqn
M = \frac{\pi}{4\Gamma^{(5)}_N} \left[r_1^2 + r_5^2 + \frac{r_0^2
ch(2\sigma)}{2}\right] = \nonumber\\ = \frac{1}{g_s^2} \left[R_9g_s
Q_1 + R_9Vg_sQ_5 + \frac{g_s^2 N}{R_9} + \frac{VR_9 r_0^2
e^{-2\sigma}}{2} \right] \label{gas'}
\eeqn
Its excitations are approximately described by transverse
oscillations generated by open strings attached to the D-branes.
These oscillations carry the momentum $N$ and are described by the gas
of the both left and right movers on the strings. Equating the energy
of this gas to $\frac{N}{R_9} + \frac{VR_9 r_0^2
e^{-2\sigma}}{2g_s^2}$ and its momentum to
$\frac{N}{R_9}$ we can determine the total energy carried by the
right and left movers.

   The entropy calculation proceeds as in the extreme case and one
finds a perfect agreement with the black hole answer \eq{noext} and
\eq{noext'} in the approximation \eq{apr}:

\beqn
S = \frac{2\pi^2 r_1r_5r_0
ch(\sigma)}{4\Gamma^{(5)}_N} = 2\pi\left(\sqrt{Q_1Q_5N_l} +
\sqrt{Q_1Q_5N_r} \right).\label{gas}
\eeqn
Moreover, one finds also an agreement for the decay rate of this
non-BPS state computed in the different (black hole and string)
pictures \cite{GuKlPe96,DhMaWa96}.  The basic process in the D-brane
picture is when a right moving string with some quantized momentum $P_9$
collides with a left moving one of the opposite momentum. They give a
closed string state of the energy equal twice the momentum which can escape
to infinity. If the moments are not exactly opposite the outgoing string
carries some momentum in the 9-th direction which means that it is charged
and very massive particle from the five-dimensional point of view. This last
emission is suppressed at low temperature, therefore, we will not consider
it.

    Let us sketch the derivation of the decay rate of the D-branes
\cite{DhMaWa96}.  From the action \eq{BI} one can find the vertex

\beq
A_D \sim G_{67}\diff x^6\diff x^7
\eeq
for the interaction of the string with the component $G_{67}$ of the
graviton - the state of the massless escaping closed string. This
component of the metric tensor looks as a scalar from the
five-dimensional point of view because the 6 and 7 are compact
directions.

   The decay rate is given by

\beq
d\Gamma(p,q,k) =
\frac{\Gamma^{(5)}_N(2\pi)^2}{8\pi R_9}\delta\left(p + q -
k\right) \frac{|A_D|^2}{p_0q_0k_0 V_4} \frac{V_4[d^4k]}{(2\pi)^4}
\eeq
where $p,q$ and $k$ are the wave vectors of the colliding strings and
of the scalar respectively; $p_0,q_0,k_0$ are their energy
components and $V_4$ denotes the volume of the spatial noncompact
four dimensions.  After the averaging over the canonical ensembles of
the left and right gases of the strings attached to the D-branes one finds
that the decay rate is equal to \cite{DhMaWa96}:

\beq
d\Gamma(k) = \frac{\pi^3 g_s^2}{V} Q_1Q_5 \omega
\frac{1}{e^{\frac{k_0}{2T_l}} - 1}
\frac{1}{e^{\frac{k_0}{2T_r}} - 1} \frac{d^4k}{(2\pi)^4}
\label{emiss'}
\eeq
where

\beqn
\frac{1}{T} = \frac12 \left(\frac{1}{T_l} + \frac{1}{T_r}\right),
\nonumber\\
T_l = \frac{1}{\pi} \frac{r_0 e^{\sigma}}{2r_1r_5}, \quad
T_r = \frac{1}{\pi} \frac{r_0 e^{-\sigma}}{2r_1r_5}
\eeqn
are the "effective" temperatures in the almost non-interacting left
and right sectors respectively \cite{DhMaWa96}. They, being equal to
$T^{-1}_{l,r} = T^{-1}(1\pm \mu)$, are some natural combinations of
the temperature and the chemical potential $\mu$, which gives the gas
some net momentum.

   Obtained answer for the decay rate should be compared with that
from the eq. \eq{emiss} for the five-dimensional black hole.
For this reason one should calculate the graybody factor
$\sigma_{gb}(k_0)$.  This calculation proceeds as follows
\cite{MaSt96}. One considers the five-dimensional Klein-Gordon
equation with the Laplacean defined through the metric in \eq{five}.
One looks for the absorption cross section of the barrier created by
this curved background. This is just what we call $\sigma_{gb}(k_0)$.
The answer is the following

\beq
\sigma_{gb} = \pi^3 r_1^2 r_5^2 \omega \frac{e^{\frac{k_0}{T}}-1}
{\left(e^{\frac{k_0}{2T_l}}-1 \right)
\left(e^{\frac{k_0}{2T_r}}-1 \right)}
\eeq
which gives the perfect agreement between the variant of the
formula \eq{emiss} in five dimensions and \eq{emiss'}.

\section*{5.4. The four-dimensional black hole}

    In a similar way as we did for the five-dimensional black hole,
one can construct the four-dimensional one. Now we
are going to work in the Type IIA rather than Type IIB theory.
For in this case it is easier to construct wanted four-dimensional
black hole.

  The solution, which we are interested in, consists of a
configuration of the $R-\tilde{R}$ 2-, 6-branes and momentum that we
had in the $d=5$ case and putting all this on the $T^6$
\cite{MaSt96'}.  In the extreme limit we obtain the black hole
solution which preserves $\frac18$ of the SUSY.  However, this black
hole has a singular geometry at the horizon. The reason is that some
of the scalar fields are unbalanced, for example, we can see from
\eq{sol} that the dilaton field will not go to a constant as we
approach the horizon, $e^{-2\varphi} = f_2^{-\frac12} f_6^{\frac32}$.

   It is interesting that one can put an additional type of charge
without breaking any additional SUSY. This charge has to be the
solitonic NS 5-brane, it is the only one allowed by the SUSY
\cite{MaSt96'}.  It also has the virtue of balancing all scalars,
for example, the dilaton now behaves as $e^{-2\varphi} =
f_2^{-\frac12} f_6^{\frac32} f_{s5}^{-1}$. In order to be more
precise let us say that our torus is $T^6 = T^4\times S'_1\times S_1$
and we have the D6-brane wrapping over all $T^6$, D2-brane wrapping
over the $S'_1\times S_1$ (directions 4,9), solitonic 5-branes
wrapping over the $T^4\times S_1$ (directions 5,6,7,8,9) and momentum
flowing along the $S_1$ (direction 9).

  Let us start with presenting the solution for the non-extreme
black hole, i.e. the "bound state" of the both $R-\tilde{R}$ branes
and anti-branes.  After doing the dimensional reduction to four
dimensions, the Einstein frame  metric (see the Appendix A) reads:

\beqn
ds^2 = - f^{-\frac12}(r)\left(1 - \frac{r_0}{r} \right)dt^2 +
f^{\frac12}(r)\left[\left(1 - \frac{r_0}{r} \right)^{-1}dr^2 + r^2
\left(d\theta^2 + sin^2(\theta) d\phi^2\right)\right] \nonumber\\
f(r) = \left(1 + \frac{r_0 sh^2(\alpha_2)}{r} \right)
\left(1 + \frac{r_0 sh^2(\alpha_5)}{r} \right)
\left(1 + \frac{r_0 sh^2(\alpha_6)}{r} \right)
\left(1 + \frac{r_0 sh^2(\alpha_p)}{r} \right).
\eeqn
This metric is parametrized by five independent quantities
$\alpha_2, \alpha_5, \alpha_6, \alpha_p$ and $r_0$.  The event
horizon lies at $r = r_0$.  The special case
$\alpha_2=\alpha_5=\alpha_6= \alpha_p$ corresponds to the
Reissner-Nordstrom metric. Hence, we see that General Relativity
solution is among the cases studied.  The overall solution contains
three additional parameters which are related to the asymptotic
values of the three scalars. From the ten-dimensional point of view,
these are the product of the radii of $T^4$, $V = R_5R_6R_7R_8$, and
the radii of the $S^1$ and $S'^1$, $R_9$ and $R_4$, and they appear in
the quantization condition for the charges.

   There are, in addition, $U(1)$ gauge fields excited,
corresponding to the four physical charges. One is the gauge field
coming from the component $G_{\mu 9}$ of the metric, which is
responsible for the "momentum charge" $N$. Then we have the
$R-\tilde{R}$ gauge field coming from the component $A_{\mu 49}$ of
the three form potential which is responsible for the D2-brane
charge, $Q_2$.  The D6-brane charge, $Q_6$, appears as the magnetic
charge for the one form $R-\tilde{R}$ potential $A_{\mu}$, and
finally the solitonic 5-brane charge, $Q_5$, also appears as the
magnetic charge for the gauge field coming from the NS antisymmetric
tensor with one index along the direction 4, $B_{\mu 4}$. The
physical charges are expressed in terms of these quantities as:

\beqn
Q_2 = \frac{r_0 V}{g_s} sh(2\alpha_2) \nonumber\\
Q_5 = r_0 R_4 sh(2\alpha_5) \nonumber\\
Q_6 = \frac{r_0}{g_s} sh(2\alpha_6)   \nonumber\\
N = \frac{r_0 V R^2_9 R_4}{g^2_s} sh(2\alpha_p)  \nonumber\\
\eeqn
where again we have set $\alpha' = 1$ and the four-dimensional Newton
constant becomes $\Gamma^{(4)}_N = \frac{g^2_s}{8VR_4R_9}$.

   The mass of the solution is equal to

\beq
M = \frac{r_0 VR_4R_9}{g^2_s} \left( ch(2\alpha_2) + ch(2\alpha_5)
+ ch(2\alpha_6) + ch(2\alpha_p) \right)
\eeq
while the entropy is

\beq
S = \frac{A_4}{4\Gamma^{(4)}_N} = \frac{8\pi r_0^2 V R_4
R_9}{g^2_s}ch(\alpha_2) ch(\alpha_5) ch(\alpha_6) ch(\alpha_p)
\eeq
The extreme limit corresponds to the $r_0\to 0,
\alpha_i\to\pm\infty$ with $Q_i$ fixed. In this limit the entropy
becomes equal to

\beq
S = 2\pi\sqrt{Q_2Q_5Q_6N}, \label{ent11}
\eeq
which is, as \eq{ent}, is U-dual and independent of the moduli.

   Now we will use the D-brane methods to recover the entropy
\eq{ent11} (see \cite{MaSt96'}). We have already seen how one can
construct the BPS state from the D-branes. The calculation of the
degeneracy in this case is almost the same as for the
five-dimensional black hole. However, there are some new ingredients
due to the solitonic 5-brane. The 5-branes intersect two kinds of the
D-branes along the $S_1$.  Different 5-branes will be at different
positions along the $S'_1$.  The D-branes can break and the ends
separate in $T^4$ when it crosses the 5-brane \cite{StTo95} (see also
the end of the section 4.3).  Hence, the $Q_2$ toroidal D2-branes break up
into the $Q_2Q_5$ cylindrical D2-branes, each of which is bounded by
a pair of 5-branes. The momentum-carrying open strings now carry an
extra label describing which pair of the 5-branes they lie in
between. The number of species becomes $N_{B,F} = 4Q_2Q_5Q_6$. The
number of BPS saturated states of this system as a function of
$Q_2,Q_5,Q_6$ and $N$ follows from the same reasoning as in the
five-dimensional case:

\beq
S = 2\pi\sqrt{\frac{(2N_B + N_F) E R_9}{12}} = 2\pi\sqrt{Q_2Q_5Q_6N},
\eeq
which indeed reproduces the classical result.

\section*{6. Conclusions}

    Thus, we see that superstring theory is able to give an
explanation of black hole thermodynamics. It is true at least for some kind
of solutions which are close to the extremity and regular on their horizons.
For arbitrary solutions, calculations presented here are not applicable.
However, in the situation when we can apply above mentioned methods, one
finds a perfect explanation of black hole thermodynamics by the means of
superstring theory. The main tool (and the only restriction) which underlies
this explanation is the presence of the $N\ge 2$ SUSY\footnote{Even if we
worked above with the four and five-dimensional $N=8$ SUSY, everything can be
straightforwardly generalized to the five-dimensional $N=4$ case
\cite{StVa96}, four-dimensional $N=4$ case \cite{JoKhMy96} and to the
four-dimensional $N=2$ case \cite{KaLoMaSt96}.}.

   Much more remains to be understood. For example, a universal
statistical explanation of the black hole entropy remains
elusive (see, however, \cite{DaSfSk97} for the most recent attempts).
Also main problem which is left is to understand the black hole
dynamics with the SUSY breaking.

   At the end it is worth mentioning that we do not know what is the
string counterpart of the {\it baby universes} and {\it warm holes}.
Therefore, we do not know how to apply string theory methods to this
situation. See, however, in the ref. \cite{Str95'} the discussion of
the so called conifold transitions in superstring theory. They are
topology change processes or, if you will, they are string theory
analogues of the gravitational warm holes. See also \cite{BeLuSa97} on the
relation between black hole singularities and the topology change
transitions.  However, such a process describes a topology change in the
internal (compact) space rather than external (our) space-time.

    For the most recent developments within M-atrix theory
\cite{BaSu} in the Schwarzschild black hole physics see \cite{Matr}.

\section*{Acknowledgments}

Author is indebted for valuable discussions to J.~Maldacena, L.B.~Okun
V.~Shadura, \\ N.A.~Voronov, A.~Zelnikov and especially to A.~Losev and
A.~Morozov.  This work was partially supported by RFFI 97-02-17927
and INTAS 96-538.

\section*{\bf Appendix A. (Elements of String Theory)}

     In this appendix we give some basic definitions for objects
appearing in superstring theory, which are needed for our
discussion in the text.  For a review of superstring theory see
\cite{GrScWi}.

   We start with the bosonic string. The generation function of
correlators in string theory is defined as follows:

\beq
Z = \sum^{\infty}_{g=0} Z_g = \sum^{\infty}_{g=0} \int [\cD
h_{ab}]_g \cD x_{\mu} e^{- i S(x_{\mu}, h_{ab})}. \label{str}
\eeq
Here the sum is over the {\it genus} $g$ of the {\it string
world-sheet} -- two-dimensional surface spaned inside a {\it target
space} by the string during its time evolution.  This sum is an
expansion over the string loop corrections.  If one considers closed
strings then these are spheres with $g$ handles, otherwise they are
discs with holes and handles of the total number $g$.
$x_{\mu}(\sigma^1,\sigma^2), \quad \mu = 0,...,d-1$ is an embedding
coordinate of the string into a $d$-dimensional target space;
$h_{ab}, \quad a,b=0,1$ is the world-sheet metric.  Also the measure
$[\cD h_{ab}]_g$ should be properly defined. Below we sketch how to do this
for the case of $g=0$.

  We proceed with the closed bosonic string. In this case the action
in \eq{str} looks as follows:

\beqn
S = \frac{1}{2\pi\alpha'}\int
d^2\sigma\Bigl\{\sqrt{h}h^{ab}G_{\mu\nu}(x)\diff_a x^{\mu}\diff_b
x^{\nu} + \nonumber\\ + \epsilon^{ab} B_{\mu\nu}(x) \diff_a
x^{\mu}\diff_b x^{\nu} + \alpha' \sqrt{h}R^{(2)} \varphi(x)\Bigr\},
\label{ac}
\eeqn
where $\sigma^a, \quad a=1,2$ are the coordinates on the string
world-sheet; $\alpha'$ is the {\it string scale} or inverse
{\it string tension}. It represents an expansion parameter of the
nonlinear $\sigma$-model \eq{ac} perturbation theory.  From the
world-sheet point of view $x_{\mu}$ are scalars enumerated by $\mu$ .
At the same time they compose the vector from the target space point
of view.

    From the world-sheet point of view the $G_{\mu\nu}$, $B_{\mu\nu}$
and $\varphi$ (the {\it graviton, antisymmetric tensor} and {\it
dilaton} respectively) represent coupling constants.  While from the
target space point of view the theory \eq{ac} represents the string
in the background field of a "gas". The latter is composed of the massless
excitations of the string (external sources in \eq{str}, \eq{ac}).
In fact, as we will show at the end of the section, the string theory has
the graviton, anti-symmetric tensor and dilaton as the massless
excitations in its spectrum. One can write down the vertex operators
for these excitations.  Then it is possible to consider the string
propagating in the flat target space (i.e. to use \eq{str} with the action
from the eq.  \eq{ac} where $G_{\mu\nu} = \eta_{\mu\nu}$, $B_{\mu\nu} = 0$
and $\varphi = 0$) and interacting with these excitations. Summing over all
tree level interactions with any number of the external legs, corresponding
to these fields, one gets exponent of the interactions. This is the action
from the eq.  \eq{ac}. To get correlation functions of these massless string
excitations, one should differentiate the eq.  \eq{str} over them.

   The vev $\varphi_0$ of the dilaton gives the coupling constant for
the mentioned genus expansion:

\beq
Z_g \sim \exp{\left\{\frac{1}{2\pi}\int d^2\sigma\sqrt{h}
R^{(2)}\varphi_0 \right\}} = e^{2(g-1)\varphi_0} = g_s^{2(g-1)},
\eeq
where $R^{(2)}$ is the two-dimensional scalar curvature. Therefore,
string theory has two expansion parameters the $g_s = e^{\varphi_0}$
and $\alpha'$.

   Two-dimensional reparametrization invariance (general covariance
on the world-sheet) of the action \eq{ac} leads to the conservation
of the world-sheet energy-momentum tensor

\beq
\diff_a T_{ab} = 0. \label{rep}
\eeq
After the gauge fixing of the reparametrization invariance one can
use new flat complex two-dimensional coordinate $z = e^{\sigma^1 +
i \sigma^2}$ and its complex conjugate.

    One also should insist on the {\it dilatational invariance}

\beq
T_{aa} = 0 \label{dil}
\eeq
required for the functional $Z$ to be well defined. Both conditions
\eq{rep} and \eq{dil} demand that the generator of the {\it conformal
transformations}\footnote{Which are holomorphic reparametrizations,
acting on the $z$.} $T(z,\bar{z}) = T_{11} - T_{22} +
2iT_{12}$ should be a {\it holomorphic} function of $z$, i.e.
$\frac{\diff}{\diff \bar{z}}T(z) = 0$. Similarly there should be an
anti-holomorphic generator $\frac{\diff}{\diff z}\bar{T}(\bar{z}) =
0$.

    The dilatational invariance of the $\sigma$-model, at the one loop
level on the world-sheet, leads to the vanishing of the
$\beta$-functions for the $G$, $B$ and $\varphi$
couplings\footnote{Briefly, the reason for this is that a non-zero
$\beta$-function leads (as in QCD) to the appearance of a mass scale
in the theory which manifestly breaks the dilatational invariance.}:

\beqn
\beta^G_{\mu\nu} = \cR_{\mu\nu} - 2\nabla_{\mu}\nabla_{\nu}\varphi +
\frac{1}{4} H_{\alpha\beta\mu} H^{\alpha\beta}_{\nu} + O(\alpha') = 0,
\qquad H_{\mu\nu\lambda} = \diff_{[\mu} B_{\nu\lambda]} \nonumber\\
\beta^B_{\mu\nu} = \nabla^{\lambda}\left(e^{2\varphi}
H_{\mu\nu\lambda}\right) + O(\alpha') = 0 \nonumber\\
\beta^{\varphi} = \frac{d - 26}{48\pi} + \alpha' \left(\cR +
\frac{1}{12} H_{\mu\nu\lambda}^2 + \left(\nabla_{\mu}\varphi\right)^2
+ 2\nabla^2_{\mu}\varphi\right) + O(\alpha'^2) = 0, \label{be}
\eeqn
where $\nabla_{\mu}$ is the covariant derivative in the metric
$G_{\mu\nu}$ and $O(\alpha')$ encodes the non-linear $\sigma$-model loop
corrections.

   Obtained equations for the string massless modes describe their
low energy dynamics. It is possible to consider these equations as
the Euler-Lagrange ones derived from the $d=26$-dimensional action:

\beq
S_{eff} \sim \int
d^{26}x\sqrt{-G}e^{-2\varphi}\left[\cR +
4\left(\nabla_{\mu}\varphi\right)^2 - \frac{1}{3}H_{\mu\nu\lambda}^2
\right] + O(\alpha'), \label{26}
\eeq
This action is written in the so called {\it string units} or
{\it string frame} or by the use of the {\it string metric}.  From
the string metric one can pass to the ordinary {\it Einstein metric
(frame)} trough the rescaling:  $G_E = G e^{4\varphi}$.  Hence, one
can see that the {\bf $26$-dimensional Einstein-Hilbert action
appears as a part of the low energy action for the massless string
modes}. It is also possible to get four dimensional gravity from string
theory via compactification \cite{GrScWi}. In this respect one says that
string theory is quantum theory of gravity.

   Let us discuss now the string spectrum. For this reason we can use the
canonical quantization of the theory \eq{ac} (defined on the
cylindrical world-sheet) in the flat target space.  General solution
of the string classical equation of motion, derived from the action
\eq{ac}\footnote{The equation takes this form only after the fixing
of the {\it conformal gauge} $h_{z\bar{z}}\sim \delta_{z\bar{z}}$ by
the use of the reparametrization invariance.},
$\diff_z\diff_{\bar{z}} x_{\mu} = 0$, in the left (depending on
$z$) sector is:

\beqn
x^{\mu}_l(z) = X_l^{\mu} - i\alpha'P^{\mu}_l \log{z} +
i\sqrt{\frac{\alpha'}{2}}\sum_{m \neq 0}\frac{1}{m}
a^{\mu}_m z^{-m}, \quad and \quad x^{\mu}(z,\bar{z}) = x^{\mu}_l(z) +
x^{\mu}_r(\bar{z}), \label{exp}
\eeqn
where $X_l^{\mu} = X_r^{\mu}$ and $P^{\mu}_l = P^{\mu}_r$ represent
the center of mass coordinate and the momentum of the string,
respectively; $a_{-m} = a^{+}_{m}$. The same expansion through
the $\tilde{a}^{\mu}_m$ is true for the case of the $x_r(\bar{z})$.

   To quantize the string, one uses the canonical commutation
relations for the left and right string modes $a^{\mu}_m$ and
$\tilde{a}^{\mu}_m$ respectively:

\beqn
[a^{\mu}_m, a^{\nu}_n] = [\tilde{a}^{\mu}_m, \tilde{a}^{\nu}_n] = i m
\delta_{m+n}\eta^{\mu\nu}; \qquad [a^{\mu}_m, \tilde{a}^{\nu}_n] = 0;
\qquad [P^{\mu}, X^{\nu}] = i\eta^{\mu\nu}, \nonumber\\
where \quad P^{\mu} = P^{\mu}_l + P^{\mu}_r \quad and \quad X^{\nu} =
X^{\nu}_l + X^{\nu}_r
\eeqn
with the Minkovskian metric $\eta^{\mu\nu}$. After the standard
definition of the vacuum $|0,0>$ (vacuum in the both left and right
sectors) one defines the states as follows:
$a^{\mu_1}_{-n_1}...\tilde{a}^{\nu_1}_{-m_1}...|0,0>$, where $\sum
n_i = N_l$ and $\sum m_i = N_r$.

   The Fourier components of the generator of the conformal invariance
$L_{-n} = \frac{1}{2\pi i}\int \frac{d z}{z^{n-1}}T(z)$ can be
expressed through the harmonics $a^{\mu}_m$ as:

\beqn
L_0 = \frac{\alpha'}{2} P^2_{\mu} + \sum_m :a^{\mu}_{-m}
\cdot a^{\mu}_m:  \qquad L_n = \frac{1}{2} \sum_m :a^{\mu}_{n+m}
\cdot a^{\mu}_{- m}:
\eeqn
Thus, $L_0$ is the string Hamiltonian and it looks like a composition
of the oscillator Hamiltonians. The $L_{n}$ generate, after
the inclusion of the terms due to the reparamerization ghosts, the
following {\it Virasoro algebra}:

\beq
[L_n,L_m] = (n-m)L_{n+m} + \frac{d - 26}{12} (m^3 - m)\delta_{n+m,0}
\eeq
where the last term is the so called conformal anomaly. It appears
from the normal ordering of the $a_m$ and $a^+_m = a_{-m}$ and of the
corresponding ghost modes. The number $c = d - 26$ is referred to as
{\it central charge} of the conformal field theory. It counts the number of
degrees of freedom on the string world-sheet. In our case, $d$ comes from the
$x_{\mu}$ and 26 is due to the ghosts.  All that is also true for the case of
the components $\tilde{L}_n$ of the $\bar{T}(\bar{z})$ expressed through the
$\tilde{a}^{\mu}_m$ modes.

  As the result of the fixing of the reparametrization invariance one
has the Virasoro constraints \cite{GrScWi} (like the Gauss law in
gauge theories and the Wheller-DeWitt equation in gravity theory)
on physical states:

\beqn
L_n|\psi> = 0, \quad \tilde{L}_n |\psi> = 0, \quad n \ge 0
\nonumber\\  \left(L_0 - \tilde{L}_0\right)|\psi> = 0; \qquad
\left(L_0 + \tilde{L}_0 - 2\right)|\psi> = 0.
\eeqn
The last line of this formula leads to

\beqn
N_l = N_r, \qquad
\frac{\alpha'}{2} P_{\mu}^2 + N_l + N_r - 2 = 0.  \label{lev}
\eeqn
Hence, one has massless mode when $N_l=N_r=1$. This is the tensor
excitation $a^{\mu}_{-1}\tilde{a}^{\nu}_{-1}|0,0>$ of the string. The
symmetric, antisymmetric parts and the trace of this excitation
correspond to the mentioned graviton, antisymmetric tensor and
dilaton fields respectively.

     The bosonic string theory also has a tachyon: $N_l=N_r=0, \quad
M^2 = - P^2 = - \frac{4}{\alpha'}$. This is the pathological
excitation because its presence means that we have chosen a wrong
(unstable) vacuum with imaginary energy.

\section*{\bf Appendix B. (Construction of the Type II Superstring
Theories)}

  To get a sensible string theory one should add SUSY on the
two-dimensional world-sheet \cite{GrScWi}. The SUSY is
added by the aid of the anticommuting $\psi_{\mu}$ fields which are
world-sheet superpartners of the $x_{\mu}$ and by the aid of
the world-sheet metric superpartner.  From the
world-sheet point of view the fields $\psi_{\mu}$ are fermions
enumerated by $\mu$. At the same time they compose a vector from the
target space point of view.

  In this appendix we discuss how one constructs the Type II
superstring theories which we use in the main body of the text. One
considers the $N=1$ two-dimensional SUGRA.  Due to the presence
of the conformal symmetry this SUGRA symmetry is enhanced to the {\it
superconformal} one.  As we discuss below one needs to do some extra
work to get SUGRA in the target space.

   In the flat target space the superstring with a cylindrical
world-sheet is described by the action:

\beq
S = \frac{1}{4\pi\alpha'} \int d^2z \left(\diff_z x^{\mu}
\diff_{\bar{z}} x_{\mu} + \bar{\psi}^{\mu} \hat{\diff}
\psi_{\mu}\right) \label{supstr}
\eeq
where we have fixed the world-sheet metric to be $h_{z\bar{z}} =
\delta_{z\bar{z}}$ and get rid of its superpartner by the use of the
reparametrization and superconformal invariances. We did not include
the reparametrization ghost terms into this action.

    There are two types of the boundary conditions on the
left and right fermions\footnote{Appearance of the both types of
boundary conditions is demanded by the modular invariance - remnant
of the conformal invariance on the world-sheets with higher
topologies.  Such an invariance exchanges with each other the sectors in
string theory obeying these boundary conditions.} on the closed string
world-sheet:  the {\it Ramond (R)} type;

\beq
\psi_{\mu}(\sigma^2 + 2\pi) = \psi_{\mu}(\sigma^2)
\eeq
and the {\it Neveu-Schwarz (NS)} type;

\beq
\psi_{\mu}(\sigma^2 + 2\pi) = - \psi_{\mu}(\sigma^2).
\eeq
and the same for the $\tilde{\psi}_{\mu}$. Therefore, in addition to
the \eq{exp}, one also has two kinds of the mode expansion for
the solutions of the free two-dimensional Dirac equation on the
cylinder:

\beqn
\psi^{\mu}(z) = \psi_0^{\mu} + \sum_n \frac{b^{\mu}_n}{z^n}  \quad
(R), \nonumber\\ \psi^{\mu}(z) =
\sum_{n}\frac{c^{\mu}_{n+\frac12}}{z^{n+\frac12}} \quad (NS)
\eeqn
and similarly for the $\tilde{\psi}$ field.

  Quantizing the superstring theory \eq{supstr}, one imposes the
standard anti-commutation relations on these modes.  For example, the
zero modes $\psi^{\mu}_0$ generate the {\it Clifford algebra}
(algebra of the Dirac $\gamma$-matrices):

\beq
\left\{\psi^{\mu}_0, \psi^{\nu}_0\right\} = \eta^{\mu\nu}. \label{cl}
\eeq

     The superstring states are constructed by the multiplication of
some state from the left sector to a state on the same level
(because of the eq. \eq{lev}) from the right one.  Thus, one has,
depending on the relative boundary conditions in the left and right
sectors, four kinds of states:

\beqn
\begin{array}{l l}
NS-\tilde{NS} \, & NS-\tilde{R} \, \\
R-\tilde{NS} \, & R-\tilde{R} \, .
\end{array} \label{state}
\eeqn
Let us discuss the left sector (discussion of the right
one is similar). Vacuum sates are:

\beqn
\begin{array}{l l l}
 & NS & R \\
P^2_{\mu} = \frac{1}{\alpha'} & |0> & --- \\
P^2_{\mu} = 0 & c^{\mu}_{-\frac{1}{2}}|0> & |0>, \,
\psi_0^{\mu}|0>, \, \psi_0^{\mu}\psi_0^{\nu}|0>...
\end{array} \label{nsr}
\eeqn
where $|0>$, in the Ramond sector, is defined below. While $|0>$ in
the Naveu-Schwarz sector is defined as a standard vacuum for
fermions.

  There is the tachyon in this spectrum.  To get rid of it, one
should project on the eigen-states of the operator $(-1)^f$ with the
eigen-value $(-1)$, where $f$ counts the fermionic number of an
operator.  This is so called {\it GSO projection} which kills the
tachyon $|0>$ and leaves the $c^{\mu}_{-\frac12}|0>$ state in the NS
sector.  It is this GSO projection which leads, after the account of
the both left and right sectors, to the appearance of the SUSY in the
target space.  After that the anti-diagonal elements in \eq{state}
give the superpartners to the diagonal ones.

  Let us discuss what happens in the $R$ sector. For this reason we
change the basis of the zero modes $\psi^{\mu}_0$ to

\beqn
d^{\pm}_0 = \frac{1}{\sqrt{2}}\left(\psi^1_0 \mp \psi^0_0\right)
\qquad d^{\pm}_i = \frac{1}{\sqrt{2i}}\left(\psi^{2i}_0 \pm
\psi^{2i+1}_0 \right), \quad i = 1,...,4.
\eeqn
Then from \eq{cl} one gets:

\beq
\left\{d^+_I, d^-_J\right\} = \delta_{IJ}, \quad I = 0,...,4.
\eeq
These $d^{\pm}_I$ generate $2^5 = 32$ Ramond ground states
$|s> = |\pm\frac12,...,\pm\frac12>$ where:

\beq
d^-_I|-\frac12,...,-\frac12> = 0, \quad d^+_I |-\frac12,...,-\frac12>
= |...,s_I=+\frac12,...>
\eeq

   As in the bosonic string there are {\it Super-Virasoro} conditions
on the physical states of the superstring: for example,
$P_{\mu}\psi^{\mu}_0|0> = 0$.  So in the frame where $P^{\mu} = (p^0,
p^0,0,...,0)$ one has that $P_{\mu}\psi^{\mu}_0 = \sqrt{2}p^0 d^+_0$
and, then, $s_0 = +\frac12$.  This leaves only $s_i = \pm \frac12,
\quad i = 1,...,4$, i.e. $16$ physical vacua: $8_s$ with the even
number of $\left(-\frac12\right)$ and $8_c$ with the odd number of
$\left(-\frac12\right)$. These $8_s$ and $8_c$ give different
chirality spinor representations of the ten-dimensional Lorentz
group.  The GSO projection keeps one among these states and removes
the other.  At the same time one has an arbitrary choice for the
vacuum:

\beq
(-1)^f|-\frac12,...,-\frac12> = \pm |-\frac12,...,-\frac12>.
\eeq
Therefore, if one chooses the opposite signs for the vacua of the
$R$ and $\tilde{R}$ sectors, one gets the non-chiral Type IIA
theory.  If the same, then one gets the chiral Type IIB.

    Thus, in the $R-\tilde{R}$ sector one has the bosonic tensor
fields $A_{\mu_1...\mu_n}$ which are related to the states obtained
as a products of that from \eq{nsr} in the left and right
sectors.  The vertex for an emission of such a string state is
defined as:

\beqn
\hat{V} \sim Q^{\alpha}
\left[\cC\hat{F}\right]_{\alpha\beta}\tilde{Q}^{\beta}, \quad \hat{F}
= F_{\mu_1...\mu_n} \gamma^{[\mu_1}...\gamma^{\mu_n]}, \nonumber\\
F_{\mu_1...\mu_n} = \diff_{[\mu_n} A_{\mu_1...\mu_{n-1}]}, \quad
\cC\gamma_{\mu}\cC^{-1} = - \gamma^T_{\mu}, \label{r-r}
\eeqn
where $Q^{\alpha}$ and $\tilde{Q}^{\alpha}$ are the two-dimensional
fields \cite{GrScWi} (compositions of the ghost fields and $\psi$'s)
which generate the target space SUSY; $F$ is the field strength of
the $R-\tilde{R}$ tensor potential $A$.  Due to the chirality
properties of the $Q$'s, forced by the GSO projection, in the Type
IIA theory there are only even rank $F$ are present, while in
the Type IIB only odd rank $F$ are present.

   The discussion of the other part of the spectrum and of the low
energy actions in the Type II theories one can find in the section 4.

\section*{Appendix C. (Elements of Duality)}

  Besides SUSY there is another tool used to control the low energy
dynamics in superstring theory. This tool is {\it duality} (for the
review see \cite{rewdu}) which helps to get some insights about
strong coupling dynamics.  That is why here we explain a few facts
about duality, which are necessary for our discussion in the main
text.

  The qualitative idea of duality came from the Fourier
transformation\footnote{In the case of free theories the duality is
simply Fourier transformation acting on the functional integral.}
which exchanges slowly varying functions with fast harmonics.
In fact, very naively, duality transformations, acting on the
functional integral of a theory, exchange fast quantum fluctuations
with slowly varying semiclassical ones. Usually duality
transformations are accompanied by the inversions of some coupling
constants.  Therefore, we can pass, via duality, from the week
coupling region to the strong one. Thus, it is possible to find what
are the dynamical degrees of freedom when microscopic ones are strongly
coupled.  Under such duality transformations some theories are
self-dual while other are exchanged with each other. In the latter case some
theory describes the strong coupling dynamics of another one.

   In superstring theory theorists use several, related to each
other, types of dualities.  These are {\it T-duality, S-duality,
U-duality} and {\it Mirror symmetry}\footnote{In this paper we will not
discuss the Mirror symmetry which is a kind of a generalization of
T-duality.}. Where, U-duality is the composition of the T- and
S-dualities when they do not commute \cite{WiHuTo95}.

   Let us begin with T-duality and then follow with S-duality. For
the review of T-duality see \cite{GiPoRa94}, while we discuss it
very qualitatively. We start with the closed bosonic string theory
which is self-dual under T-duality.  The zero modes in the expansion
\eq{exp} from the Appendix A are:

\beq
x^{\mu}(z,\bar{z}) = x^{\mu}_l(z) + x^{\mu}_r(\bar{z}) \sim -i
\sqrt{\frac{\alpha'}{2}} \left(a^{\mu}_0 +
\tilde{a}^{\mu}_0\right)\sigma^2 + \sqrt{\frac{\alpha'}{2}}\left(a^{\mu}_0
- \tilde{a}^{\mu}_0\right)\sigma^1
\eeq
where the momentum $P_{\mu}$ in the eq. \eq{exp} is equal to

\beq
P^{\mu} = P^{\mu}_l + P^{\mu}_r =
\frac{1}{\sqrt{2\alpha'}}\left(a^{\mu}_0 + \tilde{a}^{\mu}_0\right).
\eeq
Under the transformation $\sigma^1\sim\sigma^1 + 2\pi$,
$x^{\mu}(z,\bar{z})$ changes by
$2\pi\sqrt{\frac{\alpha'}{2}}\left(a^{\mu}_0 - \tilde{a}^{\mu}_0
\right)$. Because for a non-compact spatial direction $\mu$,
$x^{\mu}(z,\bar{z})$ is single valued, one has the equality

\beq
a^{\mu}_0 = \tilde{a}^{\mu}_0 = \sqrt{\frac{\alpha'}{2}} P_{\mu}.
\eeq
However, for a compact direction, say $j$, of a radius $R$, $X^j$ has
the period $2\pi R$. Then, under the transformation
$\sigma^1\sim\sigma^1 + 2\pi$, $X^j(z,\bar{z})$ can change by $2\pi m
R$, where $m$ counts the number of times our string wraps around the
compact direction. At the same time, the momentum $P^j$ can take the
values $\frac{n}{R}$. This means that

\beqn
a^j_0 + \tilde{a}^j_0 = \frac{2n}{R}\sqrt{\frac{\alpha'}{2}}
\nonumber\\
a^j_0 - \tilde{a}^j_0 = m R \sqrt{\frac{2}{\alpha'}}
\eeqn
and so

\beqn
a^j_0 = \left(\frac{n}{R} +
\frac{mR}{\alpha'}\right)\sqrt{\frac{\alpha'}{2}} \nonumber\\
\tilde{a}^j_0 =  \left(\frac{n}{R} - \frac{mR}{\alpha'}\right)
\sqrt{\frac{\alpha'}{2}}
\eeqn
Turning to the mass spectrum, we have

\beqn
M^2 = - \left(P_{\mu}\right)^2 = \frac{2}{\alpha'}
\left(a^j_0\right)^2 + \frac{4}{\alpha'} \left(N_l - 1\right) =
\left(\frac{n}{R} + \frac{mR}{\alpha'}\right)^2 + \frac{4}{\alpha'}
\left(N_l - 1\right)
\nonumber\\
= \frac{2}{\alpha'} \left(\tilde{a}^j_0\right)^2 + \frac{4}{\alpha'}
\left(N_r - 1\right) =  \left(\frac{n}{R} -
\frac{mR}{\alpha'}\right)^2 + \frac{4}{\alpha'} \left(N_r - 1\right)
\eeqn
As one can see from this formula, the first condition from the eq.
\eq{lev} is spoiled. Therefore, along the compact
directions (in contrast to the non-compact ones), strings might have
only either left or right moving modes, which we use in the section
five.

   The mass spectra of the theories at the radius equal to $R$ and
$\frac{\alpha'}{R}$ are identical if we make the exchange $n\to m$.  This
indicates that the theories at $R$ and at $\frac{1}{R}$ are identical.  Such
a flipping of $R$ is referred to as T-duality transformation.
Necessity for having winding modes $m$ for T-duality says that it is
a transformation peculiar only for extended objects which can wrap
around compact directions.

  There is a more rigorous proof that T-duality is an exact symmetry of the
bosonic string theory. It is valid for any $\alpha'$ at each order of the
expansion over $g_s$ \cite{GiPoRa94}. For example, one can show that the
partition function \eq{str} and correlators for the string are invariant
under such a transformation.  Thus, T-duality is the symmetry of the conformal
non-linear $\sigma$-model.

   T-duality exchanges the strong and the week coupling regions.
In fact, if we take the string theory on a target space with a
sufficiently large radius $R$ then the non-linear sigma model
perturbation theory (over $\alpha'$) is valid. Hence, our ordinary
space-time geometry receives only small corrections. However, as
$R\to 0$, this perturbation theory spoils and one should use
T-duality to pass to that of the dual theory at
$R' \sim \frac{1}{R}$.  In this situation $\frac{R^2}{\alpha'}$
plays the role of the coupling constant in the theory.

    In the modern language T-duality concept can be reformulated as
follows \cite{GiPoRa94}. One has the moduli space of conformal
theories parametrized by the $R$. This is the space of the unitary
(after account of the BRST symmetry) theories with the conformal
central charge (do not mix with the SUSY one) $c = d - 26$. The latter is
equal to zero in our case.  It happens that in this moduli space
there are theories which at the classical level (their actions) look
different\footnote{The last fact is due to that $G_{jj}$ in the eq.
\eq{ac} is either $\sim R$ or $\sim\frac{1}{R}$.} but their quantum
theories (correlation functions) are identical.  They are related to
each other through the exchange $R\to\frac{1}{R}$ and the momentum
modes are exchanged with the winding ones.  Moreover, one can see
that the momentum modes are charged with respect to the gauge field $G_{\mu
j}$ (such as in \eq{abel}). The charge being equal to the momentum in the
compact direction. While the winding modes are charged with respect to the
gauge field $B_{\mu j}$. The charge being equal to the winding number.
Therefore, the $G_{\mu j}$ and $B_{\mu j}$ are exchanged under T-duality
transformations \cite{GiPoRa94}.

   All that can be straightforwardly generalized to the case of the
compactification on a torus $T^n, \quad n \le 10$ which has
different radii in different directions $R_k, \quad k = 1,...,n$
\cite{GiPoRa94}.  T-duality identifies all theories which are
obtained from the one on $T^n$ with $R_k$ by flipping any number of
the $R_k$'s.  The closed bosonic string theory is self-dual under
such transformations.

   In the case of the superstring theories the situation is slightly
more complicated \cite{GiPoRa94}. For example, under T-duality
transformations the Type IIA and IIB theories are exchanged with each
other. The reason for this is that under T-duality transformations
the supercurrent in the right sector flips its chirality \cite{Pol}.
Therefore, one passes from the chiral Type IIB theory to the
non-chiral Type IIA one or vise versa.

    Let us proceed with S-duality which is more complicated than
T-duality.  In fact, T-duality, being non-perturbative from the
point of view of the $\alpha'$ corrections, is perturbative from the
point of view of the $g_s$ corrections.  Moreover, the flat ($T^n$)
compactifications do not receive quantum corrections.  Therefore, one
can find more or less rigorous proof of T-duality.  While in
favor of S-duality we have just evidences supported by consistency
checks \cite{rewdu}.  At this point the main difficulty is due to the
fact that S-duality is non-perturbative from the point of view of the
$g_s$ corrections (it exchanges $g_s\to\frac{1}{g_s}$). While one has
no other than the perturbative definition of superstring theory.

    All of the evidences in favor of S-duality are coming from the
non-renormalization theorems for the low energy SUGRA actions and for
the BPS states. The point is that, while there are {\bf no
perturbative} quantum corrections which deform these low energy SUGRA
actions, there might be {\bf non-perturbative} quantum effects
which {\bf do} change them. One can "control" these changes, looking
at different BPS states in these theories which are various p-brane
solitons.

   Thus, comparing BPS spectra and looking for the transformations
of the low energy actions (in the string frame) under the exchange
$g_s\to\frac{1}{g_s}$, one can find different relations between
all superstring theories \cite{rewdu,WiHuTo95}. Mainly this is due to
the fact that in ten dimensions, because of different anomalies, only
five superstring theories\footnote{At the same time there are plenty
of theories which can exist in $d\le 9$ dimensions.}
can exist \cite{GrScWi}.  Also, all these superstring theories
are related to M-theory \cite{rewdu,WiHuTo95}.  At low energies
this theory is described by the 11-dimensional SUGRA one.

   For example, the Type IIB theory is self-dual under S-duality
\cite{WiHuTo95}:  one can unify the dilaton and the axion fields of
this theory into a complex field $\rho = \chi + i e^{\varphi}$. It
appears that the low energy SUGRA theory is invariant under the
$SL(2,{\bf Z})$ transformations acting rationally on the $\rho$.  At
the same time the $NS-\tilde{NS}$ field $B_{\mu\nu}$ and
the $R-\tilde{R}$ field $A_{\mu\nu}$ are mixed under such
transformations. One can unite them into the two-dimensional
representation of the $SL(2,{\bf Z})$.  Therefore, under such a
duality transformation the fundamental string soliton (charged with
respect to the $B_{\mu\nu}$ field) exchanged with the $R-\tilde{R}$
1-brane (charged with respect to the $A_{\mu\nu}$ field) or with
"dyonic" strings, carrying some amounts of both types of charges.

   At the same time, the Type IIA theory in the strong coupling
($g_s\to\infty$) limit is described by M-theory \cite{WiHuTo95}.
Here eleventh dimension is compactified on a circle of the radius $R\sim
g_s^{\frac43}$ which tends to $\infty$ as $g_s\to\infty$. Therefore,
in the strong coupling region the Type IIA theory acquires extra
spatial dimension. At the same time, the F-string soliton becomes the
2-brane wrapped over the extra compact direction of the M-theory.

   Thus, one can see that S-duality exchanges fundamental
excitations with solitonic ones and also mixes corresponding tensor
gauge fields. Presently one can not prove these facts but can make different
consistency checks of these duality conjectures \cite{rewdu}.

   If one considers the compactification of the Type II theories on
the $T^n$ with $n\ge 2$ then the T- and S-dualities do not
commute\cite{WiHuTo95}.  They compose bigger U-duality. It exchanges
different gauge fields which are remnants (in the sense of \eq{abel})
of the metric, antisymmetric tensor and different $R-\tilde{R}$
fields.

    Now it is believed \cite{WiHuTo95} that all known
superstring theories are related to each other and to the
11-dimensional SUGRA through different kinds of dualities.  Hence, we
have only one theory, usually referred to as M-theory,
which is defined on a big moduli space of different parameters. Then
if we go to an infinity in any direction in the moduli space, one
gets either already known superstring theories and their
compactifications or SUGRA theories in different dimensions.



\end{document}